\documentclass[aps,prx,twocolumn,superscriptaddress,citeautoscript]{revtex4-1}

\usepackage{amsmath,amssymb}
\usepackage{bm}
\usepackage{graphicx}
\usepackage{epstopdf}
\usepackage{latexsym}
\usepackage{subfigure}
\usepackage{color}
\usepackage{dsfont} %for ensemble alphabet, and works with numbers too, use \mathds{1}
\usepackage{wasysym} %for hexagon (in Eq environment or in text)

\newcommand{\be}{\begin{equation}}
\newcommand{\ee}{\end{equation}}
\newcommand{\bea}{\begin{eqnarray}}
\newcommand{\eea}{\end{eqnarray}}

\begin{document}

\title{Definitive Evidence for Order-by-Quantum-Disorder in Er$_2$Ti$_2$O$_7$}

\author{Lucile Savary}
\affiliation{Department of Physics, University of California, Santa Barbara, CA 93106-9530, U.S.A.}
\author{Kate A. Ross}
\affiliation{Department of Physics and Astronomy, McMaster University, Hamilton, Ontario, L8S 4M1, Canada}
\author{Bruce D. Gaulin}
\affiliation{Department of Physics and Astronomy, McMaster University, Hamilton, Ontario, L8S 4M1, Canada}
\affiliation{Canadian Institute for Advanced Research, 180 Dundas St.\ W., Toronto, Ontario, M5G 1Z8, Canada}
\affiliation{Brockhouse Institute for Materials Research, McMaster University, Hamilton, Ontario, L8S 4M1, Canada}
\author{Jacob P. C. Ruff}
\affiliation{Department of Physics and Astronomy, McMaster University, Hamilton, Ontario, L8S 4M1, Canada}
\affiliation{The Advanced Photon Source, Argonne National Laboratory, Argonne, Illinois 60439, U.S.A.}
\author{Leon Balents}
\affiliation{Kavli Institute for Theoretical Physics, University of
  California, Santa Barbara, CA, 93106-4030, U.S.A.}

\date{\today}
\begin{abstract}
  Here we establish the systematic existence of a $U(1)$ degeneracy of {\sl all} symmetry-allowed Hamiltonians quadratic in the spins on the pyrochlore lattice, at the mean-field level.  By extracting the Hamiltonian of Er$_2$Ti$_2$O$_7$ from inelastic neutron scattering measurements, we then show that the $U(1)$-degenerate states of Er$_2$Ti$_2$O$_7$ are its classical ground states, and {\sl unambiguously} show that quantum fluctuations break the degeneracy in a way which is confirmed by experiment.  This is the first definitive observation of order by disorder in any material.  We provide further verifiable consequences of this phenomenon, and several additional comparisons between theory and experiment.
%Besides the precise determination of the Hamiltonian of Er$_2$Ti$_2$O$_7$ and the general statements regarding the order of the phase transitions, we establish two major results.  The first, {\sl completely general}, is the existence of a very robust $U(1)$ degeneracy of the most general spin model quadratic in spins for .  The second results from the first, because this degeneracy is a ground state degeneracy for a region of parameters in which Er$_2$Ti$_2$O$_7$ resides, and is strong verifiable evidence that Er$_2$Ti$_2$O$_7$ displays the to-date-elusive quantum order-by-disorder mechanism. %
\end{abstract}

\maketitle

Models with frustrated interactions often display an ``accidental'' ground state degeneracy in the classical limit.    Within mean field theory (MFT), the classical degeneracy extends to one of the free energy, even for quantum spins.   Theoretically, quantum or thermal fluctuations may lift this degeneracy and thereby select and stabilize an ordered state.  This phenomenon is called ``order-by-disorder'' (ObD) \cite{villain1980}, and has been discussed theoretically for more than 3 decades. 

While ObD could therefore be expected to arise fairly frequently, it has so far escaped indisputable experimental detection, to a large extent because of the difficulty of distinguishing fluctuation effects from those of weak interactions that explicitly break the degeneracy at the mean-field level.  Hence, to unambiguously identify ObD in a material, we need both a detailed knowledge of the material's Hamiltonian and a proof that a mean field degeneracy exists which is {\sl robust} to weak perturbations.  We provide both here for the rare earth pyrochlore Er$_2$Ti$_2$O$_7$, and confirm the ObD physics through confrontation of the theoretically-predicted order with experimental observations.

Prior work identified Er$_2$Ti$_2$O$_7$ as an ``XY'' antiferromagnet with an ordered ground state \cite{champion2003,champion2004,poole2007, ruff2008,petrenko2011,clarty2009,stasiak2011} in zero field.  ObD was actually already suggested for it \cite{champion2003,champion2004}, but based on an ad-hoc model which led to several significant conflicts with experiment, and as such Er$_2$Ti$_2$O$_7$ has been regarded as a long-standing puzzle.  Our model and theory go well beyond this early work and resolve all the prior enigmas. Relation to prior work on this material will be returned to at the end of the paper.

We proceed as follows. First, we prove that, at the mean-field level, {\sl any} symmetry-allowed Hamiltonian for {\sl any} magnetic material on the pyrochlore lattice, quadratic in the spins, possesses a $U(1)$ degeneracy, which can {\sl only} be broken by fluctuations or disorder. We next extract the parameters of the nearest-neighbor model for Er$_2$Ti$_2$O$_7$ from the fits of linear spin wave theory with single-crystal high-field inelastic neutron scattering, show that MFT describes Er$_2$Ti$_2$O$_7$ well, and that the $U(1)$ degeneracy of its model applies to its zero-field ordered phase.  We then calculate the splitting due to quantum fluctuations, and show that the selected state is compatible with zero-field measurements.  We also predict correspondingly a spin-wave gap of $\approx$~260~mK (and other effects) which may be measured in future experiments. 

{\sl General $U(1)$ degeneracy:}  We project the Hamiltonian to that of effective $S=1/2$ quantum spins describing the magnetic doublet of each rare earth ion on the pyrochlore lattice.  The most general form of $H$ involving two-spin interactions is $H=\frac{1}{2}\sum_{i,j}J_{ij}^{\mu\nu}S_i^\mu S_j^\nu$, where $S_i^\mu$ is the $\mu^{\rm th}$ component of the spin on the site $i$, in the {\sl global} $\left(\mathbf{\hat{x}},\mathbf{\hat{y}},\mathbf{\hat{z}}\right)$ basis.  It is implicit that the symmetries of the pyrochlore lattice constrain the relations between the $J_{ij}^{\mu\nu}$ \cite{jointpaper}. The mean field (variational) free energy $F_{\rm MF}=F_0+\langle H-H_0\rangle$, where $H_0$ and $F_0$ are the Hamiltonian and free energy for a fiducial system of decoupled spins with applied Zeeman fields, is
\begin{eqnarray}
\label{eq:free}
F_{\rm MF}&=&\frac{1}{2}\sum_{i,j}J_{ij}^{\mu\nu}m_i^\mu m_j^\nu+\frac{1}{\beta}\sum_i\Big[(\tfrac{1}{2}\!-\!|\mathbf{m}_i|)\ln\left(\tfrac{1}{2}\!-\!|\mathbf{m}_i|\right)\nonumber\\
&&+(\tfrac{1}{2}+|\mathbf{m}_i|)\ln\left(\tfrac{1}{2}+|\mathbf{m}_i|\right)\Big],
\end{eqnarray}
where $\beta=1/(k_B T)$, where $T$ is the temperature and $k_B$ is Boltzmann's constant, and where $\mathbf{m}_i=\langle\mathbf{S}_i\rangle$, $m_i^\mu=\langle S_i^\mu\rangle$ and thus $|{\bf m}_i| \leq 1/2$. The entropic part of the free energy, i.e. the last term of Eq.~\eqref{eq:free}, is obviously independent of the orientation of the magnetization $\mathbf{m}_i$.  Now consider the Ansatz
\begin{equation}
\mathbf{m}_j^0(\alpha)=\rho\,\mbox{Re}\left[e^{-i\alpha}\left(\mathbf{\hat{a}}_j+i\mathbf{\hat{b}}_j\right)\right],
\label{eq:ansatz}
\end{equation}
where $\rho\in[0,1/2]$, $\alpha\in[0,2\pi[$, and $\mathbf{\hat{a}}_j$ and $\mathbf{\hat{b}}_j$ are the {\sl local} $x$ and $y$ unit vectors, respectively (see Supplemental Material), which depend only upon which of the four sublattices the site resides. In words, Eq.~\eqref{eq:ansatz} describes translational invariant states (no unit cell enlargement) where all spins make the same angle with their local $x$-axis. (Note that this spin configuration carries no total net moment.)  This is the $\Gamma_5$ manifold of ground states identified in Ref.~\onlinecite{champion2003} for Er$_2$Ti$_2$O$_7$.  Now, let $\Phi=\rho\, e^{i\alpha}=\Phi_1+i\Phi_2$, $\Phi_1,\Phi_2\in\mathbb{R}$.  Up to an unimportant constant, the free energy for the Ansatz Eq.~\eqref{eq:free} as a function of $\Phi$ reads
\begin{equation}
F_{\rm MF}^0[\Phi]=a\Phi^2+ a^*(\Phi^*)^2+b|\Phi|^2,\qquad a\in\mathbb{C}, b\in\mathbb{R},
\end{equation}
since Eq.~\eqref{eq:free} is quadratic in the spins. Cubic symmetries then impose that $a=a^*=0$, so that $F_{\rm MF}^0$ depends on $|\Phi|$ only, i.e. solely on $|\mathbf{m}_i^0|$.   Indeed, under the three-fold rotation along the $111$ axis, one finds $\alpha \rightarrow \alpha+ 2\pi/3$, or
\begin{equation}
\Phi\rightarrow e^{2i\pi/3}\Phi\quad\Rightarrow\quad a=0,
\end{equation}
since $F_{\rm MF}^0$ should remain invariant under the above
transformation.  Thus, within MFT, the degeneracy is
present for arbitrary two-spin interactions.  Similar arguments show that the leading order term splitting the degeneracy in the free energy and consistent with cubic symmetry is
\begin{equation}
  \label{eq:3}
  F_6 = -c\, ( \Phi^6 + (\Phi^*)^6 ),
\end{equation}
with some real constant $c$.  Since there is no general argument to make $c$ vanish, we conclude that the $U(1)$ degeneracy is an artifact of the approximations introduced so far.  In MFT, it is, however, remarkably robust: {\sl six spin interactions} would be required to induce a term of the form of Eq.~\eqref{eq:3}.  In Er$_2$Ti$_2$O$_7$ (and indeed most other rare earth pyrochlores), this is entirely negligible \footnote{It arises only through sixth order virtual fluctuations into the lowest excited crystal field multiplet which is at $\approx74$~K, leading to an estimated sixth order coupling of order $10^{-9}$~meV.}.  This leaves only fluctuations -- i.e. ObD -- to determine the splitting coefficient $c$.  

{\sl Local minimum:}  By expanding about the degenerate states described by Eq.~\eqref{eq:ansatz}, we find that for arbitrary (symmetry preserving) exchange parameters, the states in Eq.~\eqref{eq:ansatz} are extrema of the free energy (see Supp. Mat.). Whether or not they are global minima, i.e. whether or not they constitute ground states of the problem, depends on the parameters $J_{ij}^{\mu\nu}$.  We now proceed to the extraction of the latter from experiment, and lift any potential suspense: for parameters relevant to Er$_2$Ti$_2$O$_7$, these are the lowest-energy states. 

%The electronic configuration of the Er$^{3+}$ ions is $4f^{11}$, so spin-orbit coupling is strong and the total angular momentum is half-integer, $J=15/2$, with a ground state doublet separated by $74.1$~K from the first excited one \cite{champion2003,gardner2010}, hence making an effective spin-1/2 model a good description at temperatures $T\ll74.1$~K.

{\sl Er$_2$Ti$_2$O$_7$ Hamiltonian: }  The effective $S=1/2$ description applies to Er$_2$Ti$_2$O$_7$ below about $74$~K \cite{champion2003,gardner2010}.  Nearest-neighbor exchange dominates, for which the Hamiltonian is constrained by symmetry to the form \cite{jointpaper} 
\begin{eqnarray}
  \label{eq:1}
  H & = & \sum_{\langle ij\rangle} \Big[ J_{zz} \mathsf{S}_i^z \mathsf{S}_j^z - J_{\pm}
  (\mathsf{S}_i^+ \mathsf{S}_j^- + \mathsf{S}_i^- \mathsf{S}_j^+) \nonumber \\
  && +\, J_{\pm\pm} \left[\gamma_{ij} \mathsf{S}_i^+ \mathsf{S}_j^+ + \gamma_{ij}^*
    \mathsf{S}_i^-\mathsf{S}_j^-\right] \nonumber \\
&& +\, J_{z\pm}\left[ \mathsf{S}_i^z (\zeta_{ij} \mathsf{S}_j^+ + \zeta^*_{ij} \mathsf{S}_j^-) +
  {i\leftrightarrow j}\right]\Big],
\end{eqnarray}
where the sans serif characters $\mathsf{S}_i^\mu$ denote components of the spins in the {\sl local} pyrochlore bases, where $\gamma$ is a $4\times4$ complex unimodular matrix, and $\zeta=-\gamma^*$ \cite{jointpaper}. The linear combinations relating the $J_{ij}^{\mu\nu}$'s (for nearest-neighbor $i$ and $j$) to $J_{zz}$, $J_{\pm}$, $J_{z\pm}$ and $J_{\pm\pm}$, the explicit expression of $\gamma$ and the local bases used in Eq.~\eqref{eq:1} are given in the Supp. Mat..  

%Because the ground state doublet of the Er 4f J=15/2 spins is very well separated from the first excited doublet, the above Hamiltonian describes . 

\begin{figure*}    
\includegraphics[width=.75\linewidth]{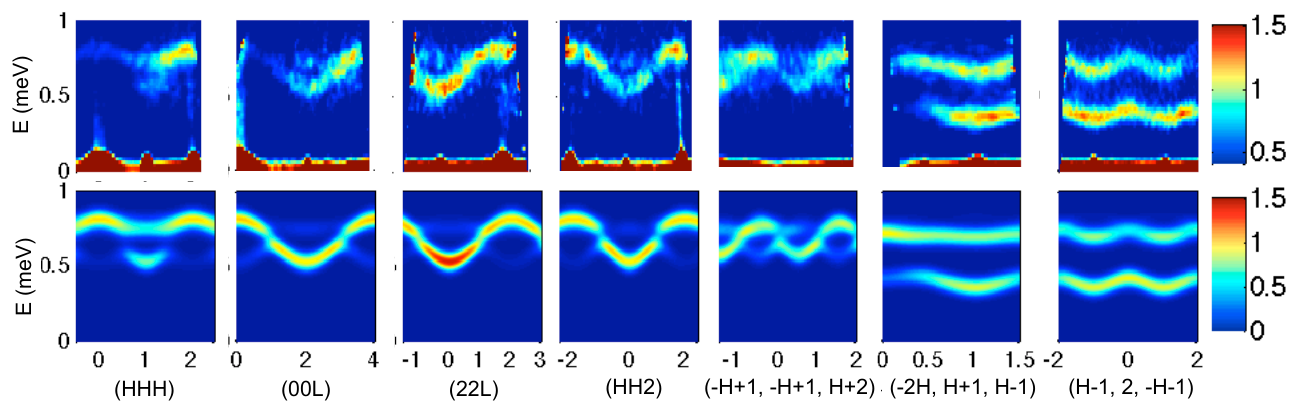}
\caption{The measured $S(\mathbf{Q},\omega)$ at $T=\;$30~mK, $H =  3$~T sliced along several directions.  The first five columns show $S(\mathbf{Q},\omega)$ in the HHL plane, with the field applied along $[1\bar{1}0]$, while the last two columns show $S(\mathbf{Q},\omega)$ for the field along $[111]$.  Top row: measured $S(\mathbf{Q},\omega)$.  Bottom row: calculated $S(\mathbf{Q},\omega)$, based on an anisotropic exchange model with six free parameters (see text) that were extracted by fitting to the measured dispersions.}
\label{fig:figure1}
\end{figure*}

%\begin{figure*} 
%   \includegraphics[width=12.9cm]{Balents_fig2.pdf} 
%   \caption{{\bf from YTO} Representations of the HHL scattering plane, showing the FCC Brillouin zone boundaries and the corresponding zone centers (labelled in terms of the conventional simple-cubic unit cell).  Blue lines indicate the directions of the five different cuts shown in Figure \ref{fig:figure1}.}
%\label{fig:figure2}
%\end{figure*}

To determine the four exchange constants and the two components of the $g$-tensor specific to Er$_2$Ti$_2$O$_7$, we fit inelastic neutron scattering data with the structure factor obtained from linear spin wave theory in high field applied to the Hamiltonian Eq.~\eqref{eq:1}.  This method was described at length in Ref.~\onlinecite{jointpaper} (esp. in its Appendix C). Experiments were carried out on a single crystal of Er$_2$Ti$_2$O$_7$ grown at McMaster University by the floating zone technique \cite{gardner1998single}.  Inelastic neutron scattering by the time-of-flight method was performed at the NIST Center for Neutron Research using the Disk Chopper Spectrometer \cite{copley2003disk}.  The incident wavelength of 5~\AA\ afforded an energy resolution of 0.09~meV.  Two orientations of the crystal were used such that the vertical axes, i.e. the crystallographic directions parallel to the applied field, were $[1\bar{1}0]$ and $[111]$.   Using two field orientations allowed an exceptionally comprehensive study of the high-field spin-wave spectra.  Furthermore, the understanding of the zero-field spectra from the ordered state was also enhanced by access to the two inequivalent scattering planes normal to the field directions.  In all color contour plots herein, the last two panels represent scattering within the plane normal to $[111]$.  All others include scattering vectors normal to $[1\bar{1}0]$.

Spin wave spectra arising in the polarized quantum paramagnetic state at
$H=3$~T and $T=30$~mK were fit to the general anisotropic exchange
model of Eq.~\eqref{eq:1} by matching the dispersions in several
directions using a least squares method.  The full
$S(\mathbf{Q},\omega)$ was not fit to the data, but followed directly
from the Hamiltonian extracted from the fit to the dispersions.  Within
the linear spin wave approximation and the nearest-neighbor model, we
find $g_z=2.45\pm0.23$ and $g_{xy}=5.97\pm0.08$
(Ref.~\onlinecite{cao2009} finds $g_z=2.6$ and $g_{xy}=6.8$), and in
$10^{-2}$ meV
\begin{eqnarray}
  \label{eq:2}
  J_{\pm\pm}=4.2\pm 0.5 ,\qquad&& J_{\pm} = 6.5 \pm 0.75,\; \\
 J_{zz} = - 2.5\pm 1.8 ,\qquad&& J_{z\pm} = - 0.88 \pm
  1.5 \;\;\;.\nonumber
\end{eqnarray}
% and a Curie-Weiss temperature (see Ref.~\onlinecite{jointpaper}, esp. Appendix B) $\Theta_{\rm CW}=-1.49$~K (Ref.~\onlinecite{blote1969} quotes $\Theta_{\rm CW}=-22$~K).  
Note that these parameters include the nearest-neighbor component of the dipolar interactions,  and that weaker further neighbor components  {\sl cannot} break the $U(1)$ degeneracy, as shown above. 

The above parameters Eq.~\eqref{eq:2} place Er$_2$Ti$_2$O$_7$ in a region of the $J_{zz}-J_\pm-J_{z\pm}-J_{\pm\pm}$ phase diagram far from spin ice.  Notably, in sharp contrast to Yb$_2$Ti$_2$O$_7$ \cite{jointpaper}, the interactions $J_\pm$ and $J_{\pm\pm}$ involving the local XY components of the spins are dominant.  Here conventional magnetic order is expected at low temperature \cite{theorypaper}, and Curie-Weiss MFT is a good starting point.  Within the latter, we obtain the $U(1)$ degenerate manifold as the zero-field ordered states.  Other predictions of MFT compare well with experiment. MFT predicts a continuous ordering transition at $T_c^{\rm MF}=2.3$~K which implies a fluctuation parameter $f=T_c^{\rm MF}/T_c\approx 2.1$, given the experimental transition temperature $T_c=1.1$~K \cite{ruff2008}.  This is much smaller than typical values of $f$ for systems with strong quantum fluctuations (c.f. $f=13$ for Yb$_2$Ti$_2$O$_7$ \cite{jointpaper}), and likely largely due to the usual {\sl thermal} fluctuation effects neglected in MFT.  The zero temperature field-induced transition (for a $\langle110\rangle$ field) with $H_c^{\rm MF}=1.74$~T, agrees perfectly with the experimental value $H_c=1.7\pm0.05$~T \footnote{$H_c$ was estimated in Ref.~\onlinecite{ruff2008} to be approximately $1.5$~T.  Our unpublished neutron scattering work involving the intensity of the (220) Bragg position, some of which is shown in Fig.~\ref{fig:braggpeaks}, identifies the transition at $H_c = 1.7 \pm 0.05$, which is consistent with the specific heat data presented in 
Ref.~\onlinecite{ruff2008}.}.

{\sl Zero-point fluctuations: }  Neglecting the tiny six spin couplings, only zero-point quantum fluctuations can break the degeneracy of a clean crystal at low temperature.  We show below that they do, though weakly, find the preferred states, and quantitatively estimate the energy splitting of the degenerate manifold.  

\begin{figure}[h]
\begin{center}
\includegraphics[width=2.55in]{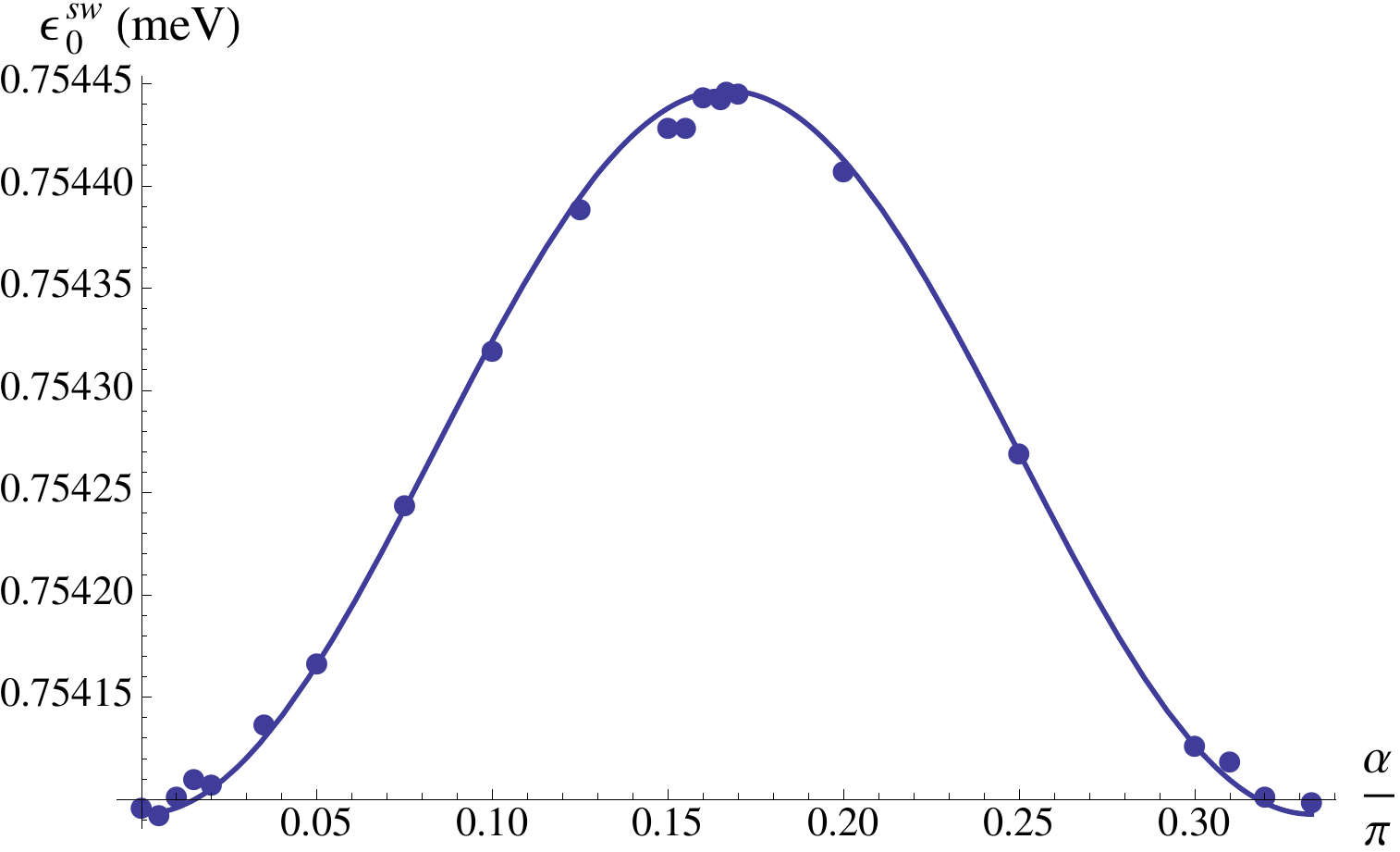}
\caption{Zero-point fluctuation energy $\epsilon^{sw}_0$ in the classically degenerate manifold parametrized by $\alpha$. The peak-to-peak energy is $\lambda\approx3.5\,10^{-4}$ meV.}
\label{fig:zeropt}
\end{center}
\end{figure}

\begin{figure}[h]
\begin{center}
\includegraphics[width=3.3in]{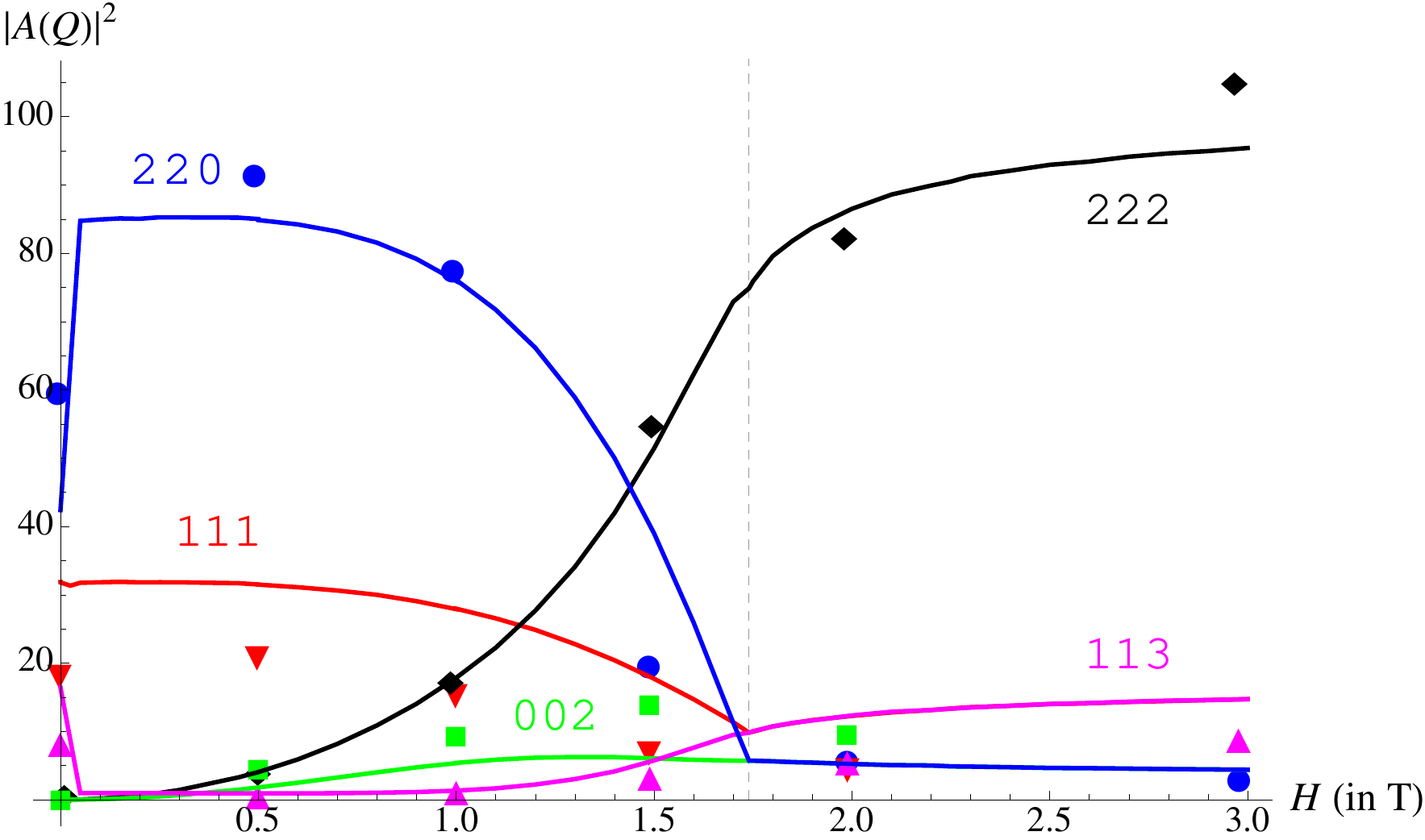}
\caption{Evolution of the Bragg peak intensities with a field
  $\mathbf{H}\parallel [\bar{1}10]$. The experimental data points from
  Ref.~\onlinecite{ruff2008} are overplotted on the theoretical curves
  (overall vertical scale of experiment was adjusted by hand) obtained when all six
  domains occupy an equal fraction of the volume in zero field.  The
  experimental values for (111) and (113) are suppressed by
  instrumental complications, which are {\sl partially} compensated for here
  by a multiplication factor of 1.3 (see Supp. Mat. for more details).  The
  dashed vertical line shows the critical field $H_c^{\rm MF}=1.74$~T
  obtained within MFT. }
\label{fig:braggpeaks}
\end{center}
\end{figure}

In the spin wave approximation, the energy of the zero-point fluctuations per unit cell is given by 
\begin{equation}
\label{eq:zeropt}
\epsilon_0^{sw}=V^{-1}_{\rm BZ}\sum_{i=1}^4\int_{\mathbf{k}\in {\rm BZ}} \omega^i_{\mathbf{k}}/2,
\end{equation}
where the sum runs over the four spin wave modes (see
Ref.~\onlinecite{jointpaper}), and where $V_{\rm BZ}$ is the volume of
the Brillouin zone.  The spectrum $\omega_\mathbf{k}^i$ for states as
described by Eq.~\eqref{eq:ansatz} depends on the angle $\alpha$ as
illustrated in the Supplemental Material, so that $\epsilon_0^{sw}$ does as well. Performing the integration in Eq.~\eqref{eq:zeropt}
numerically for different values of the phase $\alpha$, we indeed find
that zero-point fluctuations break the $U(1)$ degeneracy, and that the
six equivalent values $\alpha = n \pi/3$ ($n=0,1,\ldots,5$) are the minima of
$\epsilon_0^{sw}$ as illustrated in Figure~\ref{fig:zeropt}.  The energy
splitting fits well, up to a constant, to $\epsilon_0^{sw} = -\lambda/2
\cos 6\alpha$ ($c=32 N_{u.c.}\lambda$ in Eq.~\eqref{eq:3} at $T=0$,
where $N_{u.c.}$ is the number of unit cells), with $\lambda= 3.5\times10^{-4}$meV. 
The six $\alpha=n\pi/3$ states are equivalent, i.e. related to one
another by cubic symmetries, but differ in the absolute orientation of
the spins.  A zero-field cooled sample would be expected to form a
multi-domain state with an equal volume fraction of each state.  Indeed,
we find that an equal superposition of the spectra of all six domains
compares well with the experimental zero field neutron spectrum (see
Supp. Mat.).

{\sl Implications: } The first prediction of the ObD calculation is a
definite set of six zero-field ground states, with $\alpha = n \pi/3$,
selected by the {\sl positive} coefficient $\lambda$.  These are exactly
the $\psi_2$ states identified in Ref.~\onlinecite{champion2003}.
General symmetry arguments predict {\sl either} these $\psi_2$ states or
the alternative sequence that would be selected were $\lambda<0$, with
$\alpha = \pi/6+n\pi/3$, which are denoted $\psi_1$ states in
Ref.~\onlinecite{champion2003}.  The crucial experiment to distinguish
the two was already noted in this reference: a magnetic field applied
along $\langle110\rangle$ to a zero-field cooled sample should lead, due to domain
alignment, to a sharp {\sl increase} of the (220) Bragg peak intensity
for the $\psi_2$ states, but a sharp {\sl decrease} of intensity for the
$\psi_1$ states (see Supplemental Material). A sharp increase is
consistently observed in several experiments
\cite{champion2003,ruff2008}.  Here we make an extensive
comparison  (see Figure~\ref{fig:braggpeaks})  of theory (Supp. Mat.) to
experimental intensity versus field at {\sl
  five} Bragg peaks including (220), which gives strong evidence for the
correctness of the $\psi_2$ ground state and the Hamiltonian
parameters.  The $\psi_2$ state was also found by a sophisticated
neutron spherical polarimetry study \cite{poole2007}. 

The second consequence of our ObD scenario is the existence of a
pseudo-Goldstone mode which acquires a small gap at low temperature.  It
is important to emphasize that the exchange Hamiltonian in
Eq.~\eqref{eq:1} has only discrete (point group) symmetries, so the
appearance of a Goldstone-like mode should be surprising!  Though no
surprise seems to be expressed in the literature, the existence of such
a mode is apparent from multiple reports of a large $T^3$ low
temperature specific heat \cite{blote1969,siddharthan1999,champion2003,sosin2010,ruff2008} in
Er$_2$Ti$_2$O$_7$.  The pseudo-Goldstone mode is also explicitly visible
in our zero field inelastic neutron scattering spectra.  One can estimate the specific heat by Debye theory, $C_V^{T^3}=4N_{u.c.}\,\sigma\, T^3$, where $N_{u.c.}$ is the number of unit cells in the system, and
\begin{equation}
\sigma=\frac{k_B^4\,\pi^2\,a^3}{120\,\overline{v}^3}.
\end{equation}
Here $a$ is the usual cubic lattice spacing, and $\overline{v}$ is the
geometric mean spin wave velocity (see Supp. Mat.).  Using the
theoretical value for $\overline{v}$ one obtains $\sigma_{\rm th}
\approx 3.6\; \rm{J \cdot K}^{-4}\cdot{\rm mol}^{-1}$.  The experimental
value from Ref.~\onlinecite{ruff2008} (extracted in the Supp. Mat.) is
$\sigma_{\rm exp}=4.6$ in the same units,  comparable with theory.

Evidently the gap is not visible in current experiments.  We now estimate it
using field theory.   Consider the effective (Euclidean) action of a
system at $T=0$ with slow space and time variations of the angle $\alpha$:
\begin{equation}
\label{eq:low-en}
\mathcal{S}=
\int \frac{d^3 r}{v_{u.c.}}d\tau\left[\sum_{\mu}\frac{\kappa_\mu}{2}\left(\partial_\mu\alpha\right)^2+\frac{\eta}{2}\left(\partial_\tau\alpha\right)^2-\frac{\lambda}{2}\cos6\alpha\right],
\end{equation}
where $v_{u.c.}$ is the volume of the unit cell, and the parameters
$\kappa_\mu,\eta$ are obtained from spin wave theory (see Supp. Mat.).
Expanding the cosine above, we find that the gap
$\Delta$ to the spin waves is
\begin{equation}
\Delta=\sqrt{18\lambda/\eta}=\sqrt{27\lambda\left(J_\pm+J_{zz}/2\right)}\approx0.02\;\mbox{meV}.
\end{equation} 
This is below the 0.09~meV resolution of the inelastic neutron scattering data reported in Ref.~\onlinecite{ruff2008}, but is certainly experimentally accessible.  The gap should also manifest in a crossover from $T^3$ to activated magnetic specific heat for $T\lesssim\Delta/k_B$ (see Supp. Mat.). A nuclear Schottky anomaly below 200~mK \cite{blote1969} makes a direct observation challenging, but extrapolation of specific heat data from Ref.~\onlinecite{ruff2008} {\sl does} suggest a gap of approximately the right magnitude (Supp. Mat.).

From Eq.~\eqref{eq:low-en}, one may also extract the lengths $\xi_\mu=\sqrt{\kappa_\mu/(18\lambda)}$, which describe the width of domain walls between symmetry-related $\psi_2$ states.  We obtain $\xi_1=1.86\,a=18.71$~\AA\ and $\xi_2=2.44\,a=24.55$~\AA\ for Er$_2$Ti$_2$O$_7$.  Confrontation of domain wall theory with experiments will be addressed in a future publication.

{\sl Relation to prior theoretical work: } 
Prior theoretical work had conjectured the existence of order-by-disorder in Er$_2$Ti$_2$O$_7$, based upon a classical Heisenberg model with easy-plane single-ion anisotropy, which exhibits an {\sl extensive} degeneracy very different from the $U(1)$ degeneracy discussed here \cite{champion2003,champion2004}.  This model is microscopically inaccurate \cite{clarty2009}, and moreover the extensive degeneracy obtained within it is not robust.  The use of a general Hamiltonian, the finding of the robust degeneracy, and the extraction of the parameters of Er$_2$Ti$_2$O$_7$ are essential ingredients for the new and definitive conclusions we draw in this work.

{\sl Discussion: } The measurement of the gap via neutrons or
thermodynamics is a remaining experimental challenge, but higher
resolution experiments are needed. Neutron scattering data on
field-cooled materials which are expected to contain single domains,
i.e. single $\alpha$'s, would allow a wonderful synergy of theory and
experiment and show proof of high control on this interesting
material. The interesting field evolution of the lineshape of the Bragg reflections \cite{ruff2008} will be
returned to in a future publication. % Disorder, although probably not
% crucial in Er$_2$Ti$_2$O$_7$ since the unit cell is clearly not
% enlarged, and the Bragg peaks well resolved would be interesting to
% study, in particular since one could imagine that
% ``order-by-quenched-disorder'' could take place \cite{impuritypaper}.
% It could then compete at zero temperature with the quantum version of
% ObD described here.  A detailed study of thermal fluctuations would also
% be useful, as they could in principle lead to an interesting phase
% transition, should thermal fluctuation prefer the
% $\alpha=\pi/6\;[\pi/3]$ states. We do not {\sl a priori} suspect that
% though since no experimental signatures have been reported.  
We have achieved a conclusive and detailed understanding of the
magnetism of Er$_2$Ti$_2$O$_7$, and most importantly for the first time
shed light on a material where order-by-disorder physics is
unambiguously at play. 

After completion of this paper, a theoretical preprint \cite{zhitomirsky2012} appeared,  which reaches some of the same conclusions regarding Er$_2$Ti$_2$O$_7$.

%\begin{acknowledgments}
  We acknowledge %useful discussions with 
  Y. Qiu, K.C. Rule,
  H.A. Dabkowska, A. Bourque, and M.A. White.  K.A.R., B.D.G., and J.P.C.R. were supported by NSERC
  of Canada. L.B. and L.S. were supported by the DOE through Basic
  Energy Sciences grant DE-FG02-08ER46524, and benefitted from the
  facilities of the KITP through NSF grant PHY05-5116.
%\end{acknowledgments}

\bibliography{ETObib.bib}

\section*{\large SUPPLEMENTAL MATERIAL}

%\addtocounter{equation}{11}
%\addtocounter{figure}{3}

\section{Lattice and couplings}
\label{sec:lattice}

\subsection{Coordinates}

As usual, the coordinate system of the pyrochlore lattice is such that there is one ``up'' tetrahedron centered at the origin, with its four corners at
\begin{eqnarray}
&&\mathbf{R}_0=\frac{a}{8}\left(1,1,1\right),\quad \mathbf{R}_1=\frac{a}{8}\left(1,-1,-1\right),\\
&&\mathbf{R}_2=\frac{a}{8}\left(-1,1,-1\right),\quad\mathbf{R}_3=\frac{a}{8}\left(-1,-1,1\right),
\end{eqnarray}
where $a$ is the cubic lattice spacing (that of the underlying FCC lattice).  In Er$_2$Ti$_2$O$_7$, $a\approx 10.07$~\AA.

\subsection{Local bases}

The local cubic bases in which the Hamiltonian Eq.~\eqref{eq:1} is expressed are the following $(\mathbf{\hat{a}}_i,\mathbf{\hat{b}}_i,\mathbf{\hat{e}}_i)$ bases 
\begin{equation}
\left\{\begin{array}{l}
\mathbf{\hat{e}}_0=(1,1,1)/\sqrt{3}\\
\mathbf{\hat{e}}_1=(1,-1,-1)/\sqrt{3}\\
\mathbf{\hat{e}}_2=(-1,1,-1)/\sqrt{3}\\
\mathbf{\hat{e}}_3=(-1,-1,1)/\sqrt{3},
\end{array}\right.,
\quad
\left\{\begin{array}{l}
\mathbf{\hat{a}}_0=(-2,1,1)/\sqrt{6}\\
\mathbf{\hat{a}}_1=(-2,-1,-1)/\sqrt{6}\\
\mathbf{\hat{a}}_2=(2,1,-1)/\sqrt{6}\\
\mathbf{\hat{a}}_3=(2,-1,1)/\sqrt{6}
\end{array}\right.,
\end{equation}
$\mathbf{\hat{b}}_i=\mathbf{\hat{e}}_i\times\mathbf{\hat{a}}_i$, such that spin $\mathbf{S}_i$ on sublattice $i$ is $\mathbf{S}_i=\mathsf{S}^+_i(\mathbf{\hat{a}}_i-i\mathbf{\hat{b}}_i)/2+\mathsf{S}^-_i(\mathbf{\hat{a}}_i+i\mathbf{\hat{b}}_i)/2+\mathsf{S}^z_i\mathbf{\hat{e}}_i$.

The $4\times4$ matrix $\gamma$ introduced in Eq.~\eqref{eq:1} is 
\begin{equation}
\gamma=\begin{pmatrix}
0 & 1 & w & w^2\\
1 & 0 & w^2 & w\\
w & w^2 & 0 & 1\\
w^2 & w & 1 & 0
\end{pmatrix},
\end{equation}
where $w=e^{2\pi i/3}$ is a third root of unity.

\subsection{Relations between nearest-neighbor exchange constants}

\begin{eqnarray}
J_{zz} &=& -\frac{1}{3}(2J_1-J_2+2(J_3+2J_4)),\\
J_{\pm}&=&\frac{1}{6}(2J_1-J_2-J_3-2J_4),\\
J_{\pm\pm}&=&\frac{1}{6}(J_1+J_2-2J_3+2J_4),\\
J_{z\pm}&=&\frac{1}{3\sqrt{2}}(J_1+J_2+J_3-J_4),
\end{eqnarray}
where $J_1,..,J_4$ are the matrix elements of the exchange matrices $\mathbf{J}_{ij}$ between nearest-neighbor sites when the latter matrices are expressed in the global $(\mathbf{\hat{x}},\mathbf{\hat{y}},\mathbf{\hat{z}})$ basis. Specifically,
\begin{equation}
\mathbf{J}_{01}=
\begin{pmatrix}
J_2&J_4&J_4\\
-J_4&J_1&J_3\\
-J_4&J_3&J_1
\end{pmatrix},
\end{equation}
and the other matrices $\mathbf{J}_{ij}$ are obtained from $\mathbf{J}_{01}$ by applying the appropriate cubic rotations \cite{jointpaper}.

\subsection{Dipolar interactions}
\label{sec:dipolar}

We may very simply estimate the strength of the nearest-neighbor dipolar interactions.  We use the following notation
\begin{eqnarray}
\label{eq:dip}
H^{dip}&=&\sum_{\langle i,j\rangle}\frac{\mu_0\mu_B^2}{4\pi}\left(\frac{(\mathbf{g}_i\cdot\mathbf{S}_i)\cdot(\mathbf{g}_j\cdot\mathbf{S}_j)}{|\mathbf{r}_{ij}|^3}\right.\\
&&\left.\qquad\qquad\qquad-3\frac{(\mathbf{g}_i\cdot\mathbf{S}_i\cdot\mathbf{r}_{ij})(\mathbf{g}_j\cdot\mathbf{S}_j\cdot\mathbf{r}_{ij})}{|\mathbf{r}_{ij}|^5}\right),\nonumber
\end{eqnarray}
where $\mu_B$ is the Bohr magneton and $\mu_0=4\pi\,10^{-7}$ the universal magnetic constant. The nearest-neighbor distance is $\mathbf{r}_{\langle ij\rangle}=\sqrt{2}\,a/4$, where $a=10.07$~\AA\ is the lattice constant \cite{poole2007,gardner2010}.  Mapping Eq.~\eqref{eq:dip} to the nearest-neighbor Hamiltonian Eq.~\eqref{eq:1}, we get 
$J_{zz}^{dip} = 8.0\,10^{-3}$~meV, $J_{\pm}^{dip} = -4.6\,10^{-3}$~meV, $J_{\pm\pm}^{dip}=3.2\,10^{-2}$~meV, and $J_{z\pm}^{dip} = - 3.8\,10^{-3}$~meV, indicating that dipolar and exchange interactions are of the same order of magnitude and thus compete.  

We note, again, that further neighbor dipolar interactions will {\sl not} break the $U(1)$ degeneracy Eq.~\eqref{eq:ansatz}, as shown explicitly in the main text.

\subsection{Relation to prior model}

Previous theoretical work \cite{champion2003,champion2004} proposed the following model for Er$_2$Ti$_2$O$_7$:
\begin{equation}
H=J\sum_{\langle i,j\rangle}\mathbf{S}_i\cdot\mathbf{S}_j+D\left(\sum_i\mathbf{S}_i\cdot\mathbf{\hat{e}}_i\right)^2,\qquad J,D>0.
\end{equation}
In the $D=+\infty$ limit this model corresponds to $J_{zz}=J_{z\pm}=0$ and $J_\pm=J/6,J_{\pm\pm}=J/3$ in the language of Eq.~\eqref{eq:1}.

\section{Illustration of the $\Gamma_5$ states}
\label{sec:ansatz}

\begin{figure}[h]
\includegraphics[width=2.5in]{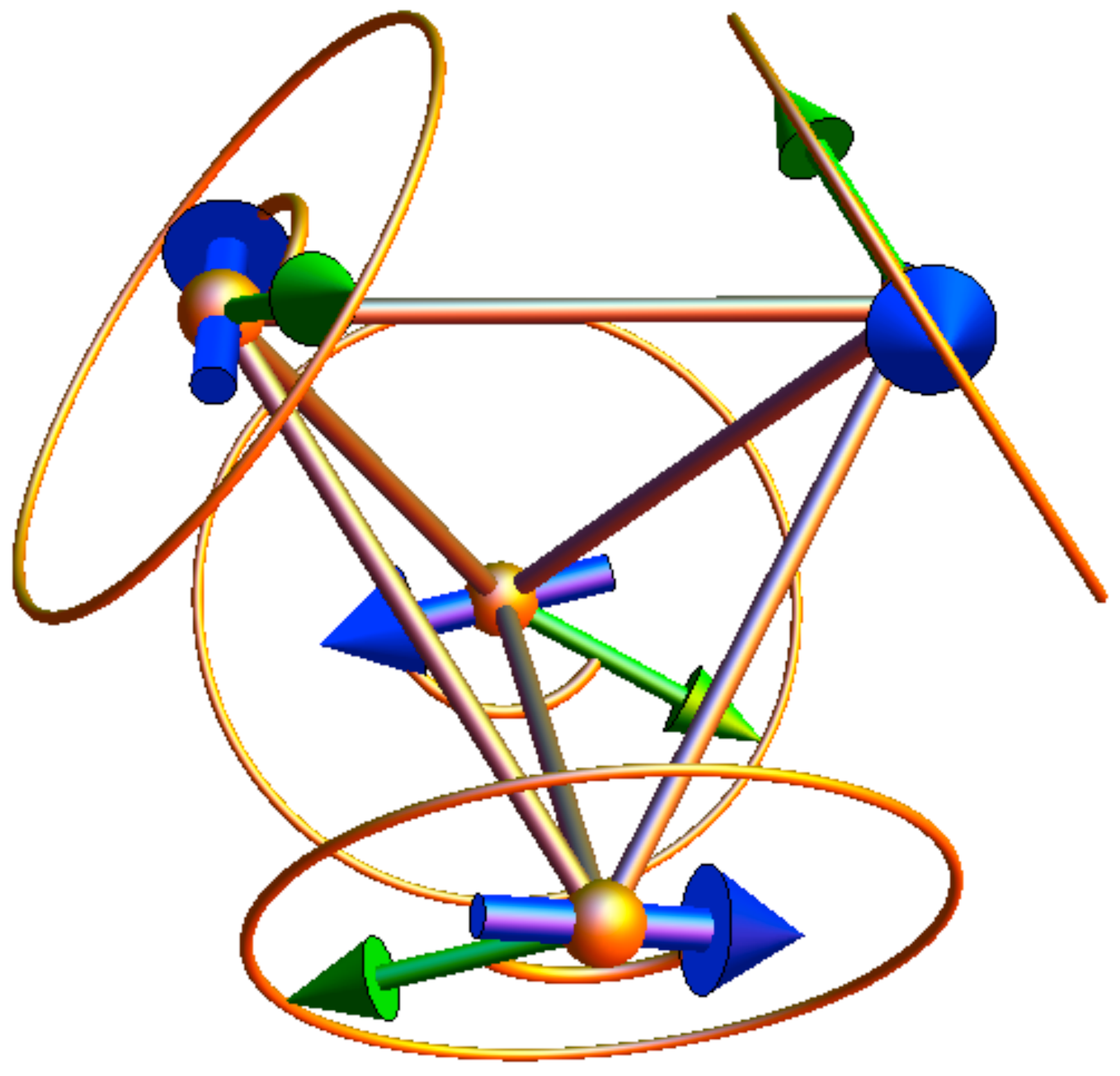} %
\caption{One of the spin states of the $U(1)$ degenerate manifold.  The
  green arrow shows the direction of the local ${\bf\hat a}_i$ axis.
  The blue arrow denotes the spin vector for one of the $\Gamma_5$
  states. }
\label{fig:angleandplane}
\end{figure}

Each state in the $\alpha=n\pi/3$ series is characterized by a global basis vector along which each of the four spin projection magnitude is largest:
\begin{equation}
\mathbf{m}_a=\pm\frac{2}{\sqrt{6}}\mathbf{\hat{x}}_\mu\pm\frac{1}{\sqrt{6}}\mathbf{\hat{x}}_{\mu+1}\pm\frac{1}{\sqrt{6}}\mathbf{\hat{x}}_{\mu+2},
\end{equation}
where
$(\mathbf{\hat{x}}_1,\mathbf{\hat{x}}_2,\mathbf{\hat{x}}_3)=(\mathbf{\hat{x}},\mathbf{\hat{y}},\mathbf{\hat{z}})$
is the usual {\sl global} basis, and $\mu$ is periodic mod 3.  For example, for $\alpha=0$, where $\mathbf{m}_a(0)=\mathbf{\hat{a}}_a$, this axis is the $\mathbf{\hat{x}}$ axis, while for $\alpha=\pi/3$, this axis is the $\mathbf{\hat{z}}$ axis:
\begin{equation}
\begin{cases}
\mathbf{m}_0(\pi/3)=(-1,-1,2)/\sqrt{6},\\
\mathbf{m}_1(\pi/3)=(-1,1,-2)/\sqrt{6},\\
\mathbf{m}_2(\pi/3)=(1,-1,-2)/\sqrt{6},\\
\mathbf{m}_3(\pi/3)=(1,1,2)/\sqrt{6}.
\end{cases}
\end{equation}
\begin{figure}[h]
\includegraphics[width=3.3in]{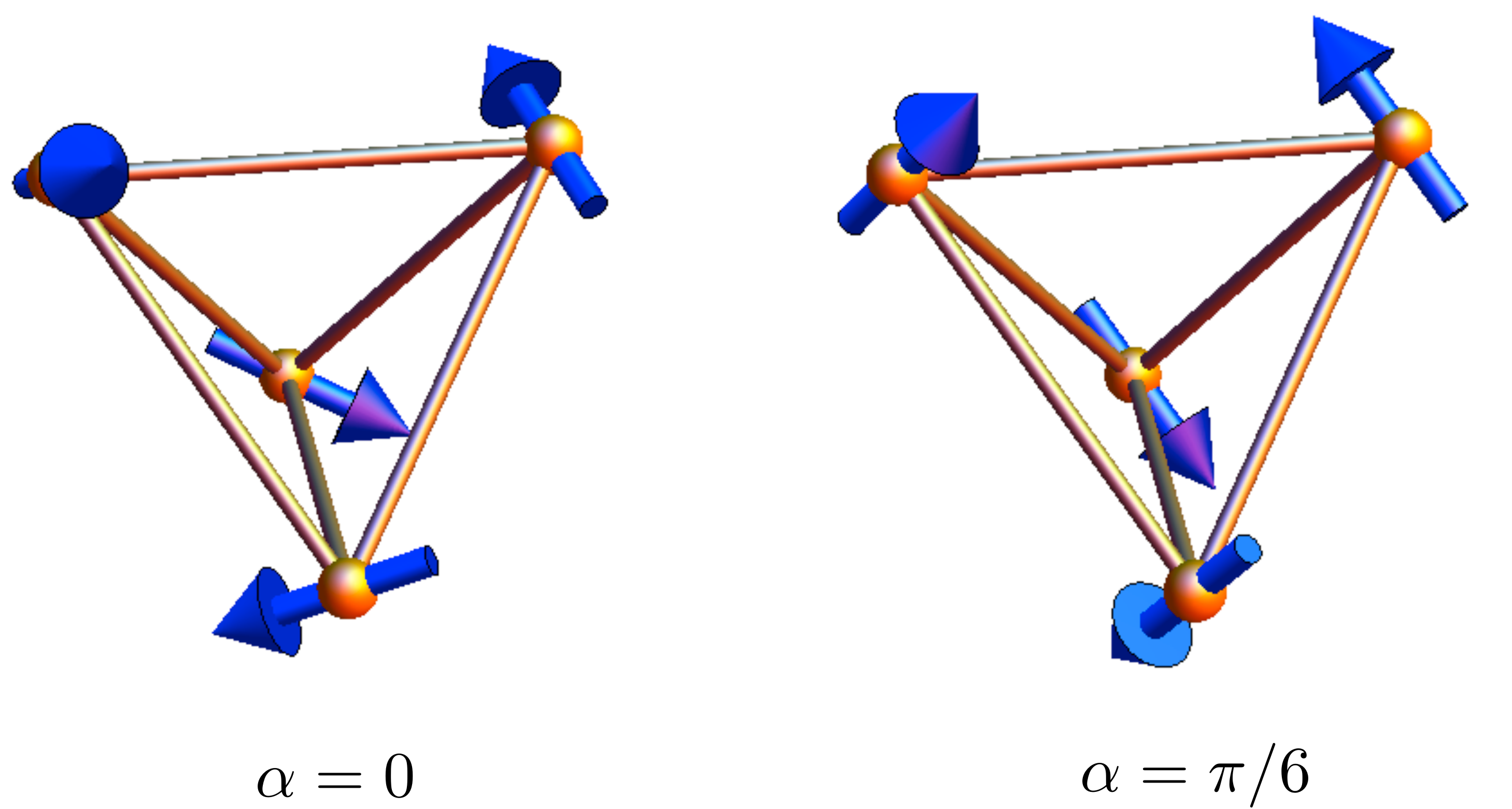} %
\caption{$\alpha=0$ and $\alpha=\pi/6$ spin states}
\label{fig:bothalphas}
\end{figure}
Each state in the $\alpha=\pi/6+n\pi/3$ series is characterized by a global basis vector along which each of the four spins have a zero projection. For example, for $\alpha=\pi/6$, this axis is the $\mathbf{\hat{y}}$ axis:
\begin{equation}
\begin{cases}
\mathbf{m}_0(\pi/6)=(-1,0,1)/\sqrt{2},\\
\mathbf{m}_1(\pi/6)=(-1,0,-1)/\sqrt{2},\\
\mathbf{m}_2(\pi/6)=(1,0,-1)/\sqrt{2},\\
\mathbf{m}_3(\pi/6)=(1,0,1)/\sqrt{2}.
\end{cases}
\end{equation}

\section{Proof of the existence of a local extremum}
\label{sec:local-extr}

%\subsection{Explicit proof}

Here we prove that the degenerate states described by Eq.~\eqref{eq:ansatz},
\begin{equation}
\mathbf{m}_j^0(\alpha)=\rho\,\mbox{Re}\left[e^{-i\alpha}\left(\mathbf{\hat{a}}_j+i\mathbf{\hat{b}}_j\right)\right],
\end{equation}
are local extrema.  To do so, we first note that, in general, for a
translationally invariant state, the spins can be written
\begin{equation}
\mathbf{m}_i=\Phi_1\mathbf{\hat{a}}_i+\Phi_2\mathbf{\hat{b}}_i+\sum_{j=1}^{10}\psi_j\mathbf{\hat{c}}_i^j,
\end{equation}
where $\Phi=\rho\, e^{i\alpha}=\Phi_1+i\Phi_2$, $\Phi_1,\Phi_2,\psi_j\in\mathbb{R}$, and where an allowed set of $\mathbf{\hat{c}}_i^j$ is such that the twelve-dimensional vectors made of the concatenation of $\{\mathbf{\hat{a}}_i\}_i$, $\{\mathbf{\hat{b}}_i\}_i$ and $\{\mathbf{\hat{c}}_i^j\}_i$ are orthogonal to one another for all $j=1,..,10$, i.e. $\begin{pmatrix}\mathbf{\hat{a}}_0^T&\cdots&\mathbf{\hat{a}}_3^T\end{pmatrix}\cdot\begin{pmatrix}\mathbf{\hat{c}}^j_0\\ \vdots\\ \mathbf{\hat{c}}_3^j\end{pmatrix}=0$, $\begin{pmatrix}\mathbf{\hat{b}}_0^T&\cdots&\mathbf{\hat{b}}_3^T\end{pmatrix}\cdot\begin{pmatrix}\mathbf{\hat{c}}^j_0\\ \vdots\\ \mathbf{\hat{c}}_3^j\end{pmatrix}=0$, and $\begin{pmatrix}({{\mathbf{\hat{c}}^l_0}})^T&\cdots&({{\mathbf{\hat{c}}^l_3}})^T\end{pmatrix}\cdot\begin{pmatrix}\mathbf{\hat{c}}^j_0\\ \vdots\\ \mathbf{\hat{c}}_3^j\end{pmatrix}=0$ for $j,l=1,..,10$ and $j\neq l$.

Now, to prove that the degenerate states constitute local extrema, we need only show that the Landau free energy around this degenerate manifold does not contain terms linear in $\Phi$ or $\psi_j$.  Terms which contain ``one'' $\Phi$ or $\psi_j$ only are readily seen to vanish because they are not invariant under time-reversal symmetry. The remaining terms (i.e. those that contain $\Phi$ {\sl and} one $\psi_j$) have the general form
\begin{equation}
\sum_{j=1}^{10}\left(f_j\,\Phi\, \psi_j+f_j^*\,\Phi^*\psi_j\right).
\end{equation}
By applying the lattice symmetry transformations to $\Phi$ and $\psi_j$, and requiring that the above term be invariant under the latter symmetries, thus imposing constraints on $f_j$, we find that $f_j=f^*_j=0$ for all $j=1,..,10$.  This concludes the proof.

%\subsection{Group theory proof}
 
%The space group of the pyrochlore lattice is $Fd\bar{3}m$, with the associated point group $O_h$.  The latter has ... irreducible representations, with dimensions ..., respectively.

\section{Fits}
\label{sec:fits}

Three-dimensional neutron scattering data sets, with two dimensions in $\mathbf{Q}$ and one in energy transfer,  were obtained by rotating the single crystal of Er$_2$Ti$_2$O$_7$ in 1.5$^\circ$ steps about the vertical axis (corresponding to the field direction, either $[1\bar{1}0]$ or $[111]$).    
Energy vs. Q slices through these three-dimensional data sets were then made in various directions in the measured $\mathbf{Q}$ plane.

\begin{figure*}[htbp]
\includegraphics[width=.9\linewidth]{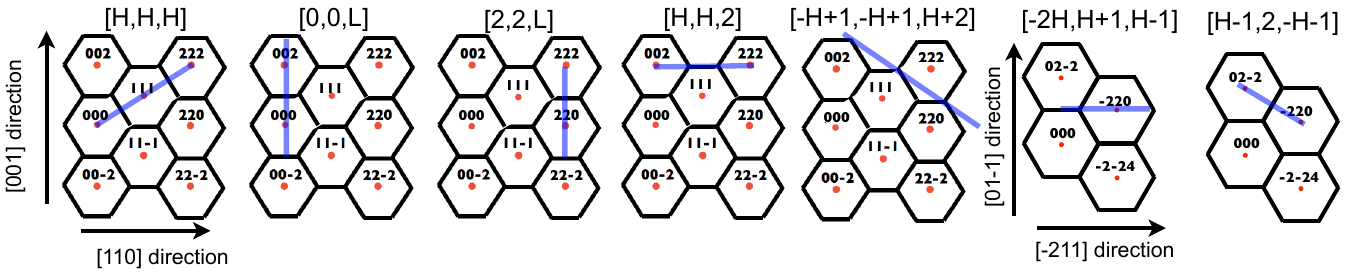} 
\caption{Representations of scattering planes perpendicular to $[1\bar{1}0]$ (i.e. the HHL plane) and $[111]$, showing the FCC Brillouin zone boundaries and the corresponding zone centers (labelled in terms of the conventional simple-cubic unit cell). Blue lines indicate the directions of the seven different cuts shown in Figures~\ref{fig:figure1}, \ref{fig:fitoverlay} and \ref{fig:superposition}.}
\label{fig:cuts}
\end{figure*}

\begin{figure*}[htbp]
\includegraphics[width=.9\linewidth]{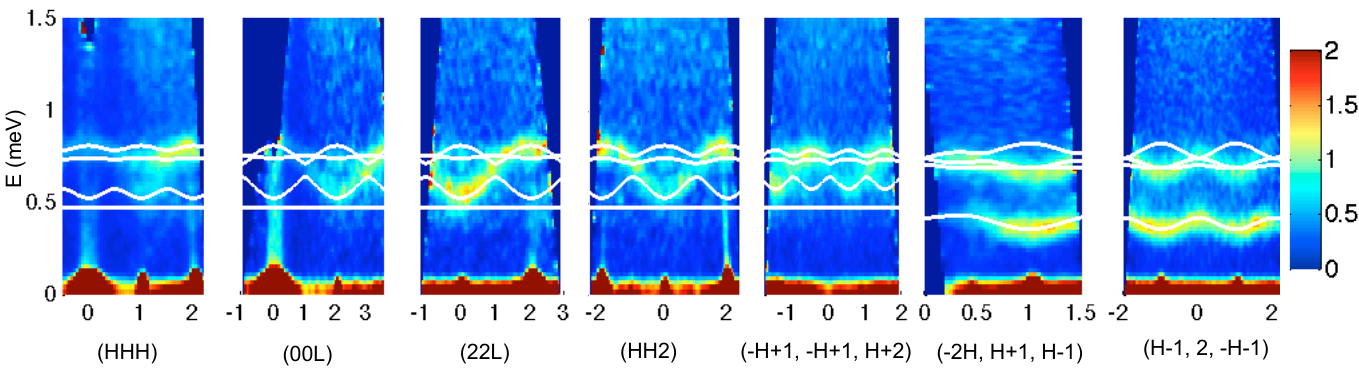} 
\caption{The dispersion curves, which were fit to the $H = 3$~T data set, are shown in white, overplotted on the data.}
\label{fig:fitoverlay}
\end{figure*}

We chose five directions for the $\mathbf{H}\|[1\bar{1}0]$ data set
and two for the $\mathbf{H}\|$[111] data set, shown in the first five
and last two columns of Figure~\ref{fig:fitoverlay}, respectively.  The cut directions are depicted in Figure~\ref{fig:cuts}.  We
performed least squares fits to the extracted spin wave dispersions in
the $H=3$~T,  $T=30$~mK data set, using theoretical values obtained
from a linear spin wave expansion of the Hamiltonian described by
Equation~\eqref{eq:1} in the main text.

Uncertainties are non-trivial to estimate in multiparameter fits.  We
proceeded by varying each exchange parameter ($J_{zz}, J_\pm$ etc.)
and $g$-factor independently, keeping the other fit parameters at their best fit
values, and assessing visually the range of acceptable fits.  These
ranges were (in meV)
\begin{eqnarray}
  \label{eq:4}
  -0.09 \leq J_{zz} \leq 0.02, \qquad 0.04 \leq J_\pm \leq 0.085,
  \nonumber\\
  0.03 \leq J_{\pm\pm}\leq 0.05, \qquad -0.012 \leq J_{z\pm} \leq 0.08,
\end{eqnarray}
while the for the $g$-factors
\begin{equation}
  \label{eq:5}
  1.8 \leq g_z \leq 3.2, \qquad 5.8 \leq g_{xy}\leq 6.3 .
\end{equation}
Notice that the range of acceptable fits is wider for $g_z, J_{zz}$
and $J_{z\pm}$ than for $J_\pm, J_{\pm\pm}$ and $g_{xy}$, which
attests to the importance of the XY spin components.  To obtain
``$\pm$'' type uncertainties as quoted in the text, we somewhat
arbitrarily took 1/3 of the half-width of the interval obtained here
for each coupling constant.  The above parameter ranges, and the best
fit values, should be viewed as a more accurate representation of the
acceptable fits.

\section{Zero-field structure factor}
\label{sec:zerofieldstructfact}

\begin{figure*}[htbp]
\includegraphics[width=.9\linewidth]{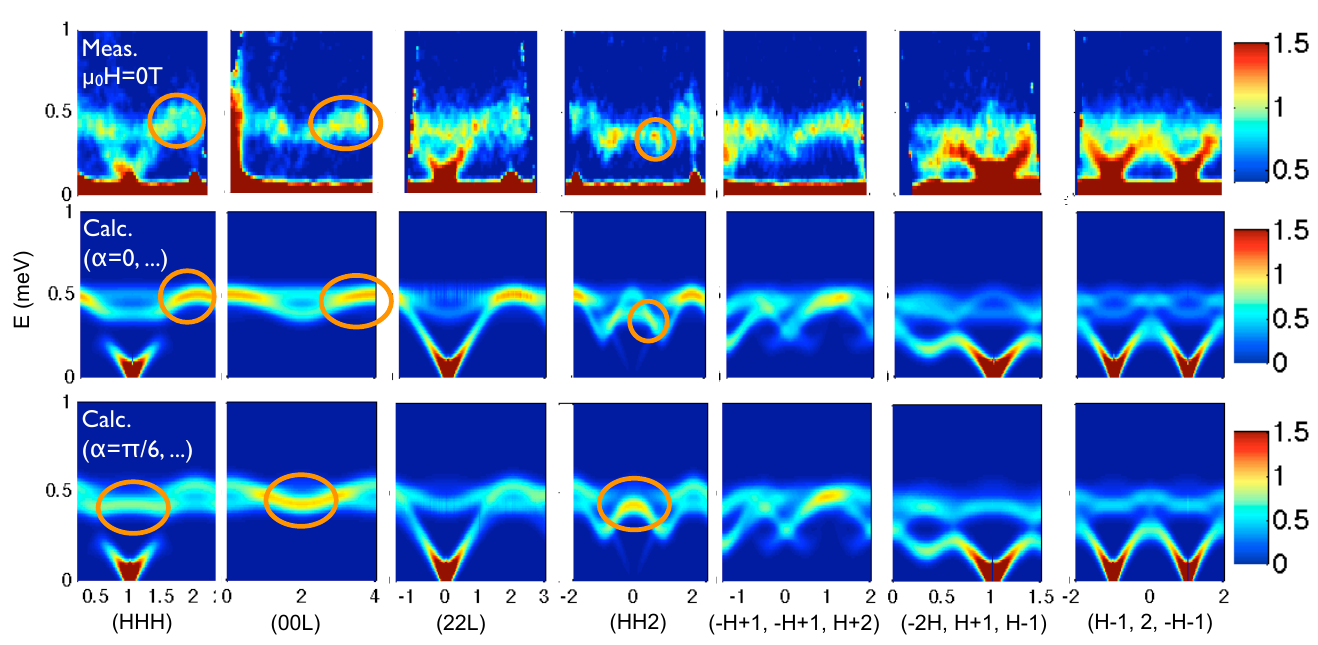} %
\caption{Comparison of the $H=0$~T, $T=30$~mK data (top row) to the calculation involving the $\alpha = n\pi/3$ (middle row) and $\alpha=\pi/6+n\pi/3$ ground states (bottom row). In each case, the domains are assumed to occupy an equal fraction of the volume. Note that the data in HHH and 00L panels has been corrected for the low $\mathbf{Q}$ self-absorption (see Fig.~\ref{fig:bragglocation}) in order to make easier visual comparisons to the calculations. Circles indicate regions where the two sets of calculations can be distinguished and compared to the data. In particular, for the HHH cut, the $\alpha=\pi/6+n\pi/3$ series shows relatively strong intensity at H=1, when the data displays weak intensity at this point.  The maximum observed around H=2 in the $\alpha=n\pi/3$ series matches the data much better.  In the 00L cut, the situation is similar: the intensity maximum at L=2 in the $\alpha=\pi/6+n\pi/3$ series disagrees with the data, which exhibits a maximum at L=4, like the spin waves of the $\alpha=n\pi/3$ series.  In the HH2 cut, the maximum of the intensity at H=0 in the $\alpha=\pi/6+n\pi/3$ series is incompatible with the data, whose intensity maximum at H$\approx$1 agrees better with the theoretical structure factor of the $\alpha=n\pi/3$ series.  We are unable to identify obvious differing features on the other cuts.}
\label{fig:superposition}
\end{figure*}

On Figure~\ref{fig:superposition}, we show the zero-field inelastic neutron scattering data, and the theoretical structure factor obtained by using the parameters fitted at 3~T, assuming that the system is made of the six equally represented symmetry-related domains $\alpha=n\pi/3$ and $\alpha=\pi/6+n\pi/3$ in rows 2 and 3, respectively.  The comparison with the $\alpha=n\pi/3$ series is very good, and we highlight a couple features which show that the structure factor obtained from the $\alpha=n\pi/3$ domains compares better than that obtained with the $\alpha=\pi/6+n\pi/3$ domains.

\section{Bragg peak intensity}
\label{sec:bragg}

\subsection{Experimental Bragg peak intensity}

Intensities for the five Bragg peak positions shown in Fig.~\ref{fig:braggpeaks} of the main text were measured at the NIST Center for Neutron Research using Disk Chopper Spectrometer using the Disk Chopper Spectrometer (DCS).   The data was obtained by rotating the crystal about the vertical $[1\bar{1}0]$ axis in 1.5$^\circ$ steps. The total intensity of each peak was summed and the nuclear contribution was subtracted using the intensities obtained at $T=2$~K, $H=0$~T.  

\begin{figure}[h]
\begin{center}
\includegraphics[width=3.3in]{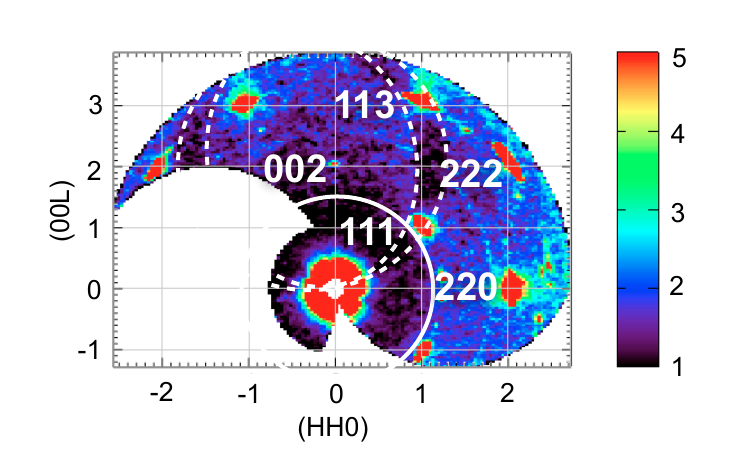}
\caption{Raw elastic scattering data at $H = 3$~T, showing the locations of the Bragg peaks.  Note that (113) and (111)  fall within the ``dark angle'' as described in Section~\ref{sec:bragg} and as denoted by dashed white lines here.  The self-absorption affecting the low $\mathbf{Q}$ region (inside the solid white line) is responsible for the unphysically low intensities near $\mathbf{Q}=\mathbf{0}$ in the (HHH) and (00L) slices shown through this paper.}
\label{fig:bragglocation}
\end{center}
\end{figure}

The field evolution of the intensity of the (220) Bragg peak has been confirmed using the FLEX triple-axis spectrometer at the Helmholtz Zentrum Berlin.  The field dependence of the peak intensity was measured at several additional field strengths, and is shown in Fig.~\ref{fig:bragglocation}.  This data agrees with that obtained from DCS for the (220) peak.

A map of the raw elastic scattering data is shown in Fig.~\ref{fig:bragglocation}.  The dashed white lines indicate the boundaries for the ``dark angle'', an area of higher neutron absorption arising from components of the magnet cryostat rotating into the incident beam.  This dark angle covers both the (111) and (113) Bragg positions, leading to artificially low integrated intensities for these peaks.  One can correct for this by measuring the reduction in incoherent elastic scattering in this area, which we have done using the high field data set in order to avoid any diffuse magnetic scattering that is present in low fields, and find a factor of $1.3\pm 0.04$ reduction in intensity.  Thus, in Fig.~\ref{fig:braggpeaks} of the main text, the intensities of the (111) and (113) peaks have been multiplied by a factor of 1.3.  It should also be noted that self-absorption effects are also present at low $\|\mathbf{Q}\|$, and begin to be important at the wave vectors indicated by the solid white line in Fig.~\ref{fig:bragglocation}.  The (111) peak thus suffers from an additional decrease in intensity, which we have not corrected for due to the difficulty in accurately doing so.  It should also be noted that the same self-absorption is responsible for the suppressed intensity near $\mathbf{Q}=0$, which is evident in all (HHH) and (00L) slices.

\subsection{Reminder of the linear spin wave theory of Ref.~\onlinecite{jointpaper}}
\label{sec:spinwaves}

In this paper, we have made extensive use of spin wave theory. It is described in detail in Appendix~C of Ref.~\onlinecite{jointpaper}, so that we use the same notations and only give here those notations and definitions needed to understand the calculations performed in this paper.

The unit vector $\mathbf{u}_a$, $a=0,..,3$ is defined to be the direction of $\mathbf{S}_a$ which minimizes the {\sl classical} energy, and $\mathbf{v}_a$ and $\mathbf{w}_a$ are such that $(\mathbf{u}_a,\mathbf{v}_a,\mathbf{w}_a)$ forms an orthonormal basis.  We further define the Holstein-Primakoff transverse bosonic operators $x_a=x_a^\dagger$ and $y_a=y_a^\dagger$ at each site such that $[x_a,y_a]=i$ and
\begin{equation}
\mathbf{S}_a\cdot\mathbf{u}_a=s-n_a,\quad\mathbf{S}_a\cdot\mathbf{v}_a=\sqrt{s}\, x_a,\quad\mathbf{S}_a\cdot\mathbf{w}_a=\sqrt{s}\,y_a,
\end{equation}
where $s=1/2$, and $n_a=\left(x_a^2+y_a^2-1\right)/2$ measures the magnetic moment due to quantum fluctuations represented by $x_a$ and $y_a$.

\subsection{Elastic structure factor at Bragg peaks}

The inelastic structure factor is proportional to
\begin{eqnarray}
&&\mathcal{I}(\mathbf{k},\omega)=\\
&&\qquad\sum_{\mu,\nu}\left[\delta_{\mu\nu}-(\mathbf{\hat{k}})_\mu(\mathbf{\hat{k}})_\nu\right]\sum_{a,b}\Big\langle M_a^\mu(-\mathbf{k},-\omega) M_b^\nu(\mathbf{k},\omega)\Big\rangle,\nonumber
\end{eqnarray}
where $M_a^\mu=\sum_\sigma g_a^{\mu\sigma} S_a^\sigma$ is the magnetic
moment operator. By definition,
\begin{equation}
M_b^\nu(\mathbf{k},\omega)=\int d\tau\, e^{i\omega\tau} e^{i\tau H}M_b^\nu(\mathbf{k})e^{-i\tau H},
\end{equation}
so, defining the above expectation value, at zero temperature,
\begin{equation}
\mathcal{M}_{ab}^{\mu\nu}(\mathbf{k},\omega)=\Big\langle0 \Big| M_a^\mu(-\mathbf{k},-\omega) M_b^\nu(\mathbf{k},\omega)\Big|0\Big\rangle,
\end{equation}
we can rewrite, using the usual spectral decomposition,
\begin{eqnarray}
&&\mathcal{M}_{ab}^{\mu\nu}(\mathbf{k},\omega)\\
&&\qquad\quad=\sum_n \delta\left(\omega+\epsilon_n-\epsilon_0\right)\langle0|M_a^\mu(-\mathbf{k}) |n\rangle\langle n|M_b^\nu(\mathbf{k})|0\rangle\nonumber.
\end{eqnarray}
Here the sum on $n$ runs on the eigenstates of $H$.  We define the amplitude
\begin{eqnarray}
\mathbf{A}_\mathbf{k}&=&\Big\langle0\Big|\sum_{a=0}^3\mathbf{M}_{a}(\mathbf{k})\Big|0\Big\rangle=\sum_{a=0}^3\mathbf{g}_a\cdot\langle0|\mathbf{S}_{a}(\mathbf{k})|0\rangle\\
&=&\sum_{a=0}^3\left(\frac{1}{2}-\langle0|n_a|0\rangle\right)\,\mathbf{g}_a\cdot\mathbf{u}_a\,e^{i\mathbf{k}\cdot\mathbf{R}_a},
\end{eqnarray}
where $\mathbf{u}_a$ is the direction of $\mathbf{S}_a$ which minimizes
the classical energy, $n_a$ is as defined in Section~\ref{sec:spinwaves}, and because $\langle x_a\rangle=\langle y_a\rangle=0$.  We obtain the elastic structure factor as the
zero frequency limit of the inelastic one (or more properly, integrating
the latter over a narrow interval of frequency near zero):
\begin{eqnarray}
&&\mathcal{I}(\mathbf{k},\omega=0)\propto\\
&&\sum_{\mu,\nu}\left[\delta_{\mu\nu}-(\mathbf{\hat{k}})_\mu(\mathbf{\hat{k}})_\nu\right]\langle0|\sum_{a} M_a^\mu(-\mathbf{k}) |0\rangle\langle 0|\sum_{b} M_b^\nu(\mathbf{k})|0\rangle\nonumber\\
&&=\mathbf{A}_{-\mathbf{k}}\cdot \mathbf{A}_{\mathbf{k}}-\left(\mathbf{\hat{k}}\cdot\mathbf{A}_{-\mathbf{k}}\right)\left(\mathbf{\hat{k}}\cdot\mathbf{A}_{\mathbf{k}}\right)\\
&&=\left|\mathbf{\hat{k}}\times\left(\mathbf{\hat{k}}\times\mathbf{A}_{\mathbf{k}}\right)\right|^2,
\end{eqnarray}
since $\mathbf{A}_{-\mathbf{k}}=\mathbf{A}_\mathbf{k}^*$.

\begin{figure}[h]
\begin{center}
\includegraphics[width=3.3in]{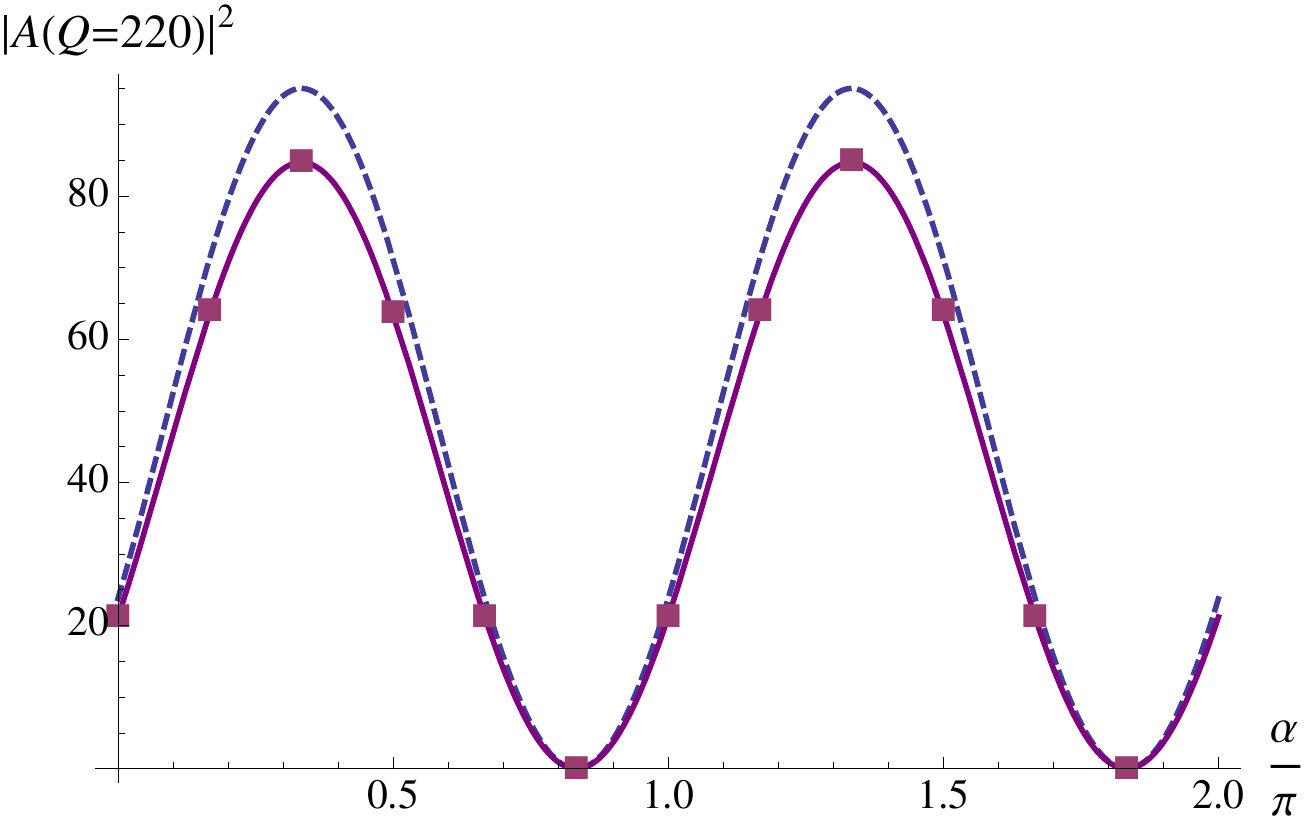}
\caption{Evolution of the $\mathbf{Q}=220$ Bragg peak intensity in zero field, as a function of $\alpha$. In blue the amplitude of the zero-temperature classical state, i.e. when $n_i=0$, with analytic formula $|\mathbf{A}_0\left(\mathbf{Q}=220,\alpha\right)|^2=\frac{8}{3}g_{xy}^2\sin^2\left[\alpha+\pi/6\right]$ is drawn.  In purple, the ``fit'' 
$|\mathbf{A}\left(\mathbf{Q}=220,\alpha\right)|^2=\frac{8}{3}g_{xy}^2\sin^2\left[\alpha+\pi/6\right]/|\mathbf{A}\left(\mathbf{Q}=220,\alpha=\pi/3\right)|^2$.  The points on the latter curve are the points with $\alpha=n\pi/6$, $n=0,..,11$.}
\label{fig:braggalpha}
\end{center}
\end{figure}

It now suffices to obtain $A_\mathbf{k}$.  However, we must account for
the domain structure.  In zero field, the six $\alpha=n\pi/3$ states are
the ground states, and we assume that they are equally present in the
system.  Then, the average intensity of a Bragg peak at $\mathbf{k}=\mathbf{Q}$ is
\begin{equation}
I_\mathbf{Q}(H=0)=\frac{1}{6}\sum_{n=0}^5|\mathbf{A}_\mathbf{Q}(n\pi/3)|^2.
\end{equation}
as shown on Figure~\ref{fig:braggalpha}.  In an applied  $[1\bar{1}0]$
field, there are only two equilibrium domains (or one above the critical
field), and both domains have equal intensity for all the Q studied, so
no averaging is necessary.  

For certain Bragg peaks, in particular
$\mathbf{Q}=\frac{2\pi}{a}\left(2,2,0\right)$, which we will also denote
$\mathbf{Q}=220$ in reciprocal lattice unit vectors, the alignment of
domains leads to a jump in intensity.  For this $\mathbf{Q}=220$ peak,
the zero field intensities of the six domains take two large and four small
values (Fig.~\ref{fig:braggalpha}), while an infinitesimal field selects
the two domains with large intensities.  This leads, ideally, to a jump
in intensity of a factor of two in passing from the six domain to two domain
state.  

We note that were we to choose the other sign of $\lambda$, such that
the $\psi_1$ states were ground states, we would obtain four large and two
small intensity contributions in zero field, and a consequent {\sl
  decrease} in intensity with applied field.  This contrasting behavior
provides a conclusive proof that the selected states are the $\psi_2$
ones ($\alpha=n\pi/3$) and not the $\psi_1$ ones
($\alpha=\pi/6+n\pi/3$).  

The domain averaging is taken in account in Fig.~3 of the main text,
which gives a similar (but opposite) effect for the $\mathbf{Q}=113$
peak, also seen experimentally.

\section{Calculation of the gap and related quantities}
\label{sec:gap}

To calculate the gap, we consider $\mathbf{H}=\mathbf{0}$ and $\alpha=0$.  Then, classically and at zero temperature, using the notations of Ref.~\onlinecite{jointpaper}, which are also given in Section~\ref{sec:spinwaves}, $2\mathbf{S}_a=\mathbf{u}_a=\mathbf{\hat{a}}_a$, and we can choose $\mathbf{v}_a=\mathbf{\hat{b}}_a$ and $\mathbf{w}_a=\mathbf{\hat{e}}_a$ to form the orthonormal basis $(\mathbf{u}_a,\mathbf{v}_a,\mathbf{w}_a)=(\mathbf{\hat{a}}_a,\mathbf{\hat{b}}_a,\mathbf{\hat{e}}_a)$. With the Ansatz Eq.~\eqref{eq:ansatz} in mind, we define 
\begin{eqnarray}
&&\Psi_0^T=\begin{pmatrix}1&1&1&1\end{pmatrix},\quad\Psi_1^T=\begin{pmatrix}1&-1&0&0\end{pmatrix},\\
&&\Psi_2^T=\begin{pmatrix}0&0&1&-1\end{pmatrix},
\quad \Psi_3^T=\begin{pmatrix}1&1&-1&-1\end{pmatrix}
\end{eqnarray}
and parameterize
\begin{equation}
X^T=\begin{pmatrix}x_0&x_1&x_2&x_3\end{pmatrix}=\alpha\,\Psi_0^T+\sum_{i=1}^3\,\chi_i\,\Psi_i^T,
\end{equation}
and use, as in Ref.~\onlinecite{jointpaper}, $Y^T=\begin{pmatrix}y_0&y_1&y_2&y_3\end{pmatrix}$. The action of the linear spin wave theory developed in Ref.~\onlinecite{jointpaper} is
\begin{equation}
\mathcal{S}=\frac{1}{2\beta}\sum_n\sum_\mathbf{k} \begin{pmatrix}X_{-\mathbf{k},-\omega_n}^T&Y_{-\mathbf{k},-\omega_n}^T\end{pmatrix}M_{\mathbf{k},\omega_n}\begin{pmatrix}X_{\mathbf{k},\omega_n}\\Y_{\mathbf{k},\omega_n}\end{pmatrix}
\end{equation}
where
\begin{equation}
M_{\mathbf{k},\omega_n}=
\begin{pmatrix}
2A_\mathbf{k}&2C_\mathbf{k}-i(i\omega_n)I_4\\
2C_\mathbf{k}^T+i(i\omega_n)I_4&2B_\mathbf{k}
\end{pmatrix},
\end{equation}
with $A_\mathbf{k}$, $B_\mathbf{k}$ and $C_\mathbf{k}$ as defined in Ref.~\onlinecite{jointpaper}. 
If we integrate out $Y$, we are left with
\begin{eqnarray}
\mathcal{S}'&=&\frac{1}{2\beta}\sum_n\sum_\mathbf{k}X_{-\mathbf{k},-\omega_n}^T\cdot N_{\mathbf{k},\omega_n}
\cdot X_{\mathbf{k},\omega_n},
\end{eqnarray}
where
\begin{equation}
N_{\mathbf{k},\omega_n}=\left[T_\mathbf{k}+i(i\omega_n)V_\mathbf{k}-(i\omega_n)^2W_\mathbf{k}\right],
\end{equation}
with
\begin{eqnarray}
T_\mathbf{k}&=&2\left(A_\mathbf{k}-C_\mathbf{k}\cdot B^{-1}_\mathbf{k}\cdot C_\mathbf{k}^T\right),\\
V_\mathbf{k}&=&\left(B^{-1}_\mathbf{k}\cdot C_\mathbf{k}^T-C_\mathbf{k}\cdot B^{-1}_\mathbf{k}\right),\\
W_\mathbf{k}&=&\frac{1}{2} B^{-1}_\mathbf{k}.
\end{eqnarray}
We want to expand $N_{\mathbf{k},\omega_n}$ to second order in $\mathbf{k}$ and $\omega_n$ about $\mathbf{k}=\mathbf{0}$ and $\omega_n=0$ (we are looking at the pseudo-Goldstone mode), so we expand $T_\mathbf{k}$, $V_\mathbf{k}$ and $W_\mathbf{k}$ to second, first and zeroth order, respectively.  Then, defining $L^{(n)}$ as the $n^{\rm th}$ order term of matrix $L$, $N_{\mathbf{k},\omega_n}$ is
\begin{equation}
N_{\mathbf{k},\omega_n}\approx\left[T^{(0)}+T_\mathbf{k}^{(2)}+i(i\omega_n)V^{(0)}-(i\omega_n)^2W^{(0)}\right]
\end{equation}
because $T_\mathbf{k}^{(1)}=V_\mathbf{k}^{(1)}=0$ since the $\mathbf{k}$ dependence comes solely from $\cos\left(\mathbf{k}\cdot(\mathbf{r}_a-\mathbf{r}_b)\right)$ terms, whose expansion involves even powers of $k^\mu$ only.  Importantly, we further find 
$\Psi_0^T\cdot V^{(0)}\cdot\Psi_j=0$ (and in fact $V^{(0)}=0$), $\Psi_0^T\cdot T^{(0)}\cdot\Psi_j=0$ and $\Psi_j^T\cdot T^{(0)}\cdot\Psi_j\neq0$ for $j=1,2,3$.  So, if we complete the squares of the $\chi_j$ terms, only the following terms involve $\alpha$ and $\chi_j$:
\begin{equation}
a^j_{\mathbf{k},\omega_n}\left(\chi_j+\frac{b^j_{\mathbf{k}}}{2\,a^j_{\mathbf{k},\omega_n}}\alpha\right)^2,
\end{equation}
with $a^j_{\mathbf{k},\omega_n}=t_j^{(0)}+\sum_{\mu,\nu}t_{j,\mu\nu}^{(2)}k_\mu k_\nu-w^{(0)}(i\omega_n)^2$, and $b_\mathbf{k}^j=\sum_{\mu,\nu}{t'_{j,\mu\nu}}^{(2)}k_\mu k_\nu-{w'}^{(0)}(i\omega_n)^2$, (with $t_j^{(0)},t_{j,\mu\nu}^{(2)},w^{(0)},{t'_{j,\mu\nu}}^{(2)},{w'}^{(0)}\in\mathbb{R}$) so that $\alpha^2$ extra terms that arise from integrating out $\chi_j$ are at least of order four in $\mathbf{k}$ and $\omega_n$.  This result is reasonable since only $\alpha$ describes the continuous degeneracy and is thus expected, alone, to give rise to the Goldstone mode.

We find, for the ground state $\alpha=0$,
\begin{eqnarray}
\mathcal{S}''&=&\frac{1}{2\beta}\sum_n\sum_\mathbf{k}\alpha_{-\mathbf{k},-\omega_n}\alpha_{\mathbf{k},\omega_n}\qquad\qquad\\
&&
\qquad\qquad\quad\times\left[\kappa_x k_x^2+\kappa_{yz}(k_y^2+k_z^2)-\eta(i\omega_n)^2 \right]\nonumber
\end{eqnarray}
where 
\begin{eqnarray}
\kappa_x&=&\frac{1}{4}\left(2J_\pm-J_{\pm\pm}\right)\\
\kappa_{yz}&=&\frac{1}{8}\left(4J_\pm+J_{\pm\pm}\right)\\
\eta=\frac{\kappa}{v^2}&=&\frac{4}{3}\frac{1}{2J_\pm+J_{zz}},
\end{eqnarray}
so that the action of the full spin-wave and zero-point fluctuation problem is
\begin{eqnarray}
\mathcal{S}'&=&\frac{1}{2\beta}\sum_n\sum_\mathbf{k}\alpha_{-\mathbf{k},-\omega_n}\alpha_{\mathbf{k},\omega_n}\\
&&\quad\quad\times\left[\kappa_x k_x^2+\kappa_{yz}(k_y^2+k_z^2)-\eta(i\omega_n)^2 +18\lambda\right],\nonumber
\end{eqnarray}
where we expanded the cosine, $-\frac{\lambda}{2}\cos6\alpha\approx-\frac{\lambda}{2}+9\lambda\alpha^2$ and took into account the overall $1/2$ prefactor.  The gap $\Delta$ is then
\begin{equation}
\Delta=\sqrt{\frac{18\lambda}{\eta}}=\sqrt{27\lambda\left(J_\pm+\frac{J_{zz}}{2}\right)}\approx 0.0222\;\mbox{meV},
\end{equation}
since $\lambda=3.51\times10^{-4}$~meV.  In Kelvins, this is $\Delta=258$ mK.
Now, if we define
\begin{equation}
v_\mu=\sqrt{\frac{\kappa_\mu}{\eta}},
\end{equation}
we get
\begin{equation}
\hat{v}_x\approx0.0413\mbox{ meV}\quad\mbox{and}\quad \hat{v}_{y,z}\approx0.0541\mbox{ meV}
\end{equation}
where $a$ is the lattice constant, and $\hat{v}_i=v_i/a$'s has the dimension of an energy.  Now, plugging in $a=10.04$ \AA, we get
\begin{equation}
v_x\approx0.416\;\mbox{meV.\AA}\qquad\mbox{and}\qquad v_{yz}\approx0.545\;\mbox{meV.\AA}.
\end{equation}
Note that the specific form of the anisotropy, i.e. $v_x\neq v_y=v_z$ is due to the choice $\alpha=0$.  The other combinations are found for other values of $\alpha=n\pi/3$.  For all the latter the velocity which we denote $v_1$ appears once, while $v_2$ appears twice.  Here $v_1=v_x$ and $v_2=v_y=v_z=v_{y,z}$. The read-off slopes of the Goldstone modes of the spin wave theory are
\begin{equation}
\hat{v}_x^{read}=0.0412\;\mbox{meV}\qquad\mbox{and}\qquad \hat{v}_{y,z}^{read}=0.0541\;\mbox{meV},
\end{equation}
i.e. a basically exact match.  We can also define two length scales,
\begin{equation}
\xi_\mu=\sqrt{\frac{\kappa_\mu}{18\lambda}},
\end{equation}
and we get
\begin{equation}
\xi_x=1.86\,a\approx 18.71\;\mbox{\AA}\quad\mbox{and}\quad\xi_{yz}=2.44\,a\approx24.55\;\mbox{\AA},
\end{equation}
where $a$ is, again, the lattice spacing.  Those length scales physically represent the lengths over which the system sees no degeneracy breaking (cf. $\lambda=0\Rightarrow\xi_i\rightarrow\infty$), and are the typical extent of domain walls in the system, if any.

\section{Specific heat}
\label{sec:specificheat}

\begin{figure}[h]
\begin{center}
\includegraphics[width=3.3in]{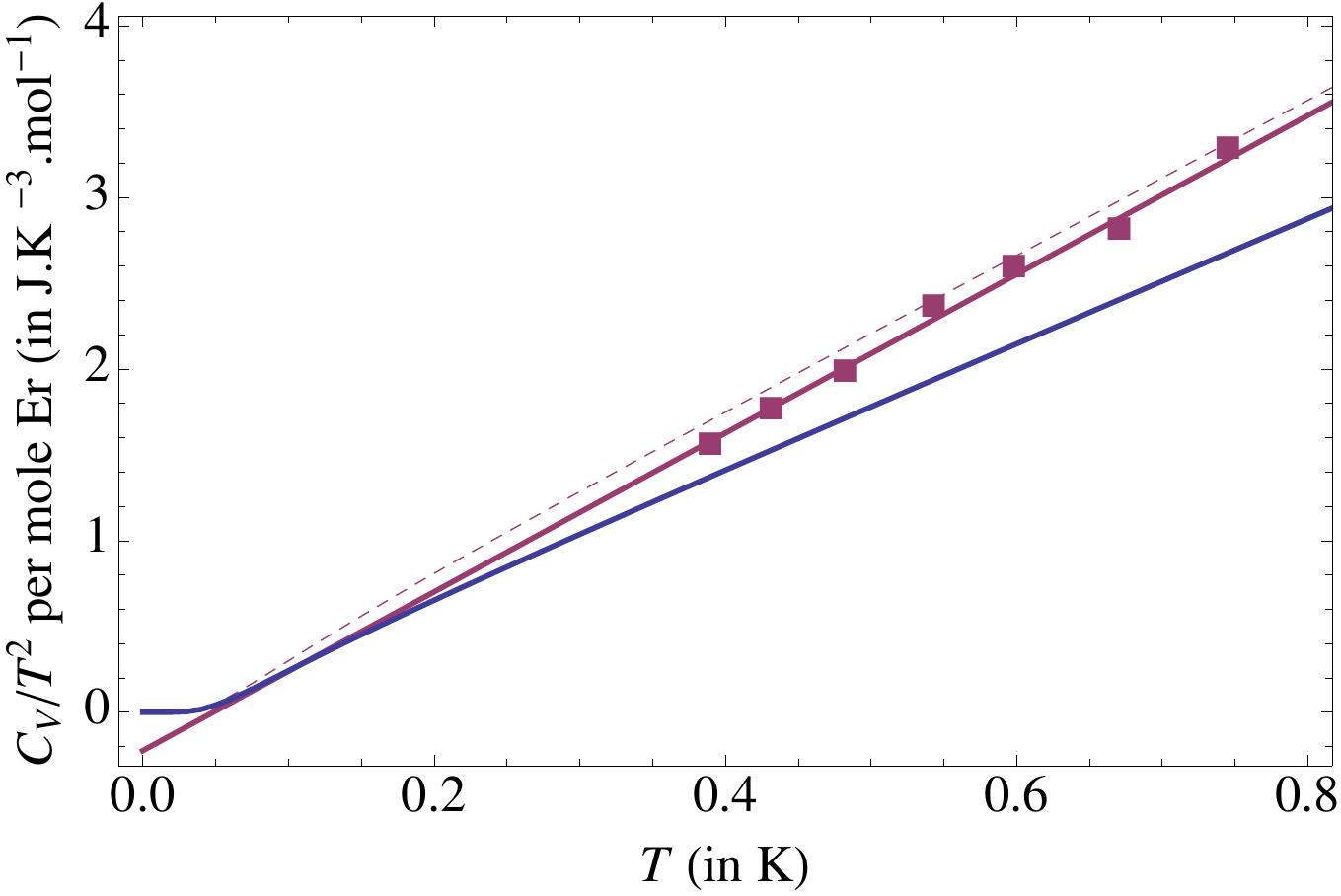}
\caption{$C_V/T^2$ versus $T$.  In blue, the theoretical plot Eq.~\eqref{eq:fullcv} obtained for the parameters of our fits in zero field, with $\Delta=0.02$~meV.  Above $0.05$~K the linear behavior of $C_V/T^2$ is clear, with a slope $\tilde{\sigma}_{\rm th}=\mathcal{N}_A\frac{k_B^4\,\pi^2\,a^3}{120\,\overline{v}^3}\approx3.62\;\mbox{J.K$^{-4}$.mol$^{-1}$}$.  The purple points are the experimental data points reported in Ref.~\onlinecite{ruff2008} (note that in the latter reference, the vertical axes is $C_p/R$ per mole of {\sl formula unit} Er$_2$Ti$_2$O$_7$), and the solid purple line is the best linear fit to the six data points with lowest $T$.  The equation of this line is $-0.2 + 4.6\,T$~J.K$^{-3}$.mol$^{-1}$ ($\tilde{\sigma}_{\rm exp}=4.6$~J.K$^{-4}$.mol$^{-1}$).  The thin dashed line's equation is $\frac{r\, \tilde{\sigma}_{\rm exp}}{16\pi^4/15}\left(\int_0^\infty dX\,\frac{X^2\left(X^2+\delta^2\right)}{\sinh^2\frac{\sqrt{X^2+\delta^2}}{2}}\right)T$, where $r$ is the ratio of $\tilde{\sigma}_{\rm th}$ to the slope of the the blue line.}
\label{fig:cv}
\end{center}
\end{figure}

\subsection{Calculation of the specific heat}

For ease of notation, we take $v_1=v_x$ and $v_2=v_y=v_z=v_{yz}$, which corresponds to the $\alpha=0$ state (as defined and mentioned in Section~\ref{sec:gap}). Of course none of the conclusions drawn here depend on this specific choice. The dispersion relation of the low-energy spin-wave mode is ($\hbar=1$)
\begin{equation}
\omega_\mathbf{k}=\epsilon_\mathbf{k}=\sqrt{v_x^2 k_x^2+v_{yz}^2\left(k_{y}^2+k_{z}^2\right)+\Delta^2},
\end{equation}
hence (the counting goes: there are as many such modes as there are unit cells), following Debye, the energy of the system is
\begin{equation}
E=\sum_\mathbf{k}\epsilon_\mathbf{k}\frac{1}{e^{\beta\epsilon_\mathbf{k}}-1}=\frac{N_{u.c.}}{V_{\rm BZ}}\int_{\rm BZ} d^3 k\, \epsilon_\mathbf{k}\frac{1}{e^{\beta\epsilon_\mathbf{k}}-1},
\end{equation}
where $V_{\rm BZ}=\frac{32\pi^3}{a^3}$ is the volume of the Brillouin zone and  $N_{\rm BZ}=N_{u.c.}$ is the number of points in the Brillouin zone (equal to the number of unit cells).  We now change the integration variables (rescale)
\begin{equation}
\tilde{k}_x= k_x v_x,\quad \tilde{k}_{y,z}= k_{y,z} v_{yz},\quad{\tilde{\mathbf{k}}}=\sum_\mu\tilde{k}_\mu\mathbf{\hat{x}}_\mu,
\end{equation}
(note $\tilde{k}_\mu$ has the units of an energy), to get
\begin{eqnarray}
E&=&\frac{1}{\overline{v}^3}\frac{N_{u.c.}}{V_{\rm BZ}}\int_{``{\rm BZ}\times\overline{v}^3"} d^3 \tilde{k}\, \tilde{\epsilon}_{\tilde{\mathbf{k}}}
\frac{1}{e^{\beta\tilde{\epsilon}_{\tilde{\mathbf{k}}}}-1}\\
&=&\frac{4\pi }{\overline{v}^3}\frac{N_{u.c.}}{V_{\rm BZ}}\int_0^\infty d\tilde{k}\, \tilde{k}^2\frac{\tilde{\epsilon}_{\tilde{\mathbf{k}}}}{e^{\beta\tilde{\epsilon}_{\tilde{\mathbf{k}}}}-1},
\end{eqnarray}
where $\tilde{\epsilon}_{\tilde{\mathbf{k}}}=\sqrt{{\tilde{\mathbf{k}}}^2+\Delta^2}$, $\overline{v}=\left(v_x v_y v_z\right)^{1/3}=\left(v_1 v_2^2\right)^{1/3}$ is the geometric mean of the velocities, where the integration runs to infinity because we have assumed $\overline{v}\Lambda\gg k_B T$, where $\Lambda$ is the ultraviolet cut-off, $\Lambda\sim1/a$.  The specific heat $C_V=\frac{\partial E}{\partial T}$ is
\begin{eqnarray}
\label{eq:fullcv}
C_V&=&\frac{4\pi }{\overline{v}^3 k_BT^2}\frac{N_{u.c.}}{V_{\rm BZ}}\int_0^\infty d\tilde{k}\, \frac{k^2\,\tilde{\epsilon}_{\tilde{\mathbf{k}}}^2\, e^{\beta\tilde{\epsilon}_{\tilde{\mathbf{k}}}}}{\left(e^{\beta\tilde{\epsilon}_{\tilde{\mathbf{k}}}}-1\right)^2}\\
&=&4N_{u.c.}\left(\frac{k_B^4 \,a^3}{128\,\pi^2\, \overline{v}^3}\int_0^\infty dX\,\frac{X^2\left(X^2+\delta^2\right)}{\sinh^2\frac{\sqrt{X^2+\delta^2}}{2}}\right)T^3,\nonumber
\end{eqnarray}
where $X=\beta \tilde{k}$ and $\delta=\beta\Delta$ are dimensionless. $C_V/(4N_{u.c.})$ is plotted in blue on Figure~\ref{fig:cv}.

\subsection{Estimate of the coefficient of the $T^3$ term}

The theoretical (blue) curve on Figure~\ref{fig:cv} is made of two parts.  Below $T\gtrsim0.05$~K the behavior is that of an activated $C_V$, while the straight line above  $T\gtrsim0.05$~K clearly pertains to the $T^3$ behavior. Setting $\Delta=0$, i.e. $\delta=0$ in Eq.~\eqref{eq:fullcv}, we extract the slope of this line:
\begin{equation}
\label{eq:ourcv1}
C_V^{\Delta=0}=4N_{u.c.}\sigma\, T^3,
\end{equation}
where
\begin{equation}
\sigma=\frac{k_B^4\,\pi^2\,a^3}{120\,\overline{v}^3}
\approx3.75\,10^{-2}\;\mbox{meV.K$^{-4}$},
\end{equation}
which means, per Er spin,
\begin{equation}
\label{eq:ourcv2}
\frac{C_V^{\Delta=0}}{{\rm Er}}\equiv\frac{C_V^{\Delta=0}}{N_{\rm Er}}=\sigma\, T^3,
\end{equation}
with the above value of $\sigma$, where $N_{\rm Er}$ is the number of Er spins or, in units more commonly used in the literature, 
\begin{equation}
\label{eq:ourcv3}
\frac{C_V^{\Delta=0}}{\rm moles\; of\; Er}=\tilde{\sigma}\, T^3,\qquad\mbox{with}\qquad \tilde{\sigma}\approx3.62\;\mbox{J.K$^{-4}$.mol$^{-1}$},
\end{equation}
since $\tilde{\sigma}=\sigma\, \mathcal{N}_A\,(\mbox{Joules per 1
  meV})$, where $\mathcal{N}_A$ is the Avogadro constant and
$\mbox{(Joules per 1 meV)}\approx1.60\,10^{-22}$ J/meV.  Note that
$\tilde{\sigma}$ is denoted $\sigma$ in the main text for cosmetic
reasons.

\end{document}